\documentclass[prd,aps,floatfix]{revtex4}

\usepackage{graphicx}
\usepackage{amsmath}
\usepackage{amssymb}
\usepackage{longtable}

\newcommand{\be}{\begin{equation}}
\newcommand{\ee}{\end{equation}}
\newcommand{\bea}{\begin{eqnarray}}
\newcommand{\eea}{\end{eqnarray}}
\newcommand{\bi}{\begin{itemize}}
\newcommand{\ei}{\end{itemize}}
\newcommand{\ben}{\begin{enumerate}}
\newcommand{\een}{\end{enumerate}}
\newcommand{\bt}{\begin{tabbing}}
\newcommand{\et}{\end{tabbing}}

\newcommand{\calO}{{\mathcal O}}

\begin{document}

%
%
\vspace*{-10mm}
\begin{flushright}
\normalsize
 KEK-CP-152        \\
 UTHEP-490         \\
 UTCCS-P-4         \\
 HUPD-0402         \\
\end{flushright}

\title{
Non-perturbative $O(a)$-improvement of Wilson quark action
in three-flavor QCD with plaquette gauge action
}

\author{
   N.~Yamada$^{1}$, 
   S.~Aoki$^2$, 
   M.~Fukugita$^{3}$, 
   S.~Hashimoto$^{1}$, 
   K-I.~Ishikawa$^{4}$,
   N.~Ishizuka$^{2,5}$, 
   Y.~Iwasaki$^{2,5}$, 
   K.~Kanaya$^{2,5}$, 
   T.~Kaneko$^{1}$, 
   Y.~Kuramashi$^{2,5}$,
   M.~Okawa$^{4}$, 
   Y.~Taniguchi$^2$,
   N.~Tsutsui$^{1}$, 
   A.~Ukawa$^{2,5}$, 
   and T.~Yoshi\'e$^{2,5}$\\
   (CP-PACS and JLQCD Collaborations)
}

\affiliation{
$^1$High Energy Accelerator Research Organization(KEK), Tsukuba, Ibaraki 305-0801, Japan \\
$^2$Institute of Physics, University of Tsukuba, Tsukuba, Ibaraki 305-8571, Japan \\
$^3$Institute for Cosmic Ray Research, University of Tokyo, Kashiwa 277-8582, Japan \\
$^4$Department of Physics, Hiroshima University, Higashi-Hiroshima, Hiroshima 739-8526, Japan\\
$^5$Center for Computational Sciences, University of Tsukuba, Tsukuba, Ibaraki 305-8577, Japan 
}

\date{\today}

\begin{abstract}
We perform a non-perturbative determination of 
the $O(a)$-improvement coefficient $c_{\rm SW}$ 
for the Wilson quark action
in three-flavor QCD with the plaquette gauge action.
Numerical simulations are carried out 
in a range of $\beta\!=\!12.0$\,--\,5.2
on a single lattice size of $8^3 \times 16$ 
employing the Schr\"odinger functional setup of lattice QCD.
As our main result, we obtain an interpolation formula
for $c_{\rm SW}$ and the critical hopping parameter 
$K_c$ as a function of the bare coupling.
This enables us to remove $O(a)$ scaling violation
from physical observables 
in future numerical simulation in the wide range of $\beta$.
Our analysis with a perturbatively modified 
improvement condition for $c_{\rm SW}$
suggests that 
finite volume effects in $c_{\rm SW}$ are not large 
on the $8^3 \times 16$ lattice.
We investigate $N_f$ dependence of $c_{\rm SW}$
by additional simulations for $N_f\!=\!4$, 2 and 0 
at $\beta\!=\!9.6$.
As a preparatory step for this study, we also determine 
$c_{\rm SW}$ in two-flavor QCD at $\beta\!=\!5.2$.
At this $\beta$, several groups carried out large-scale 
calculations of the hadron spectrum,
while no systematic determination of $c_{\rm SW}$ has been performed.

\end{abstract}


\maketitle


\section{Introduction}
\label{sec:introduction}


Recent large-scale simulations 
in two-flavor QCD have demonstrated that 
quenching artifacts found in various physical observables
are reduced by dynamical effects of up and down quarks.
%
There has been significant progress also in 
the algorithms for QCD with odd numbers of flavors:
while the conventional Hybrid Monte Carlo (HMC) algorithm~\cite{HMC}
is applicable only to even-flavor QCD,
the exact algorithms,
such as the Multi-Boson~\cite{MultiBoson} and 
the Polynomial HMC algorithms~\cite{PHMC} 
capable of odd-flavor cases,
have been developed.
Clearly the time has come to undertake
a fully realistic and extensive simulations 
of QCD with dynamical up, down and strange quarks.

%
%
Since simulations with dynamical quarks are
computationally demanding, 
highly improved lattice actions should be employed
in the three-flavor simulations.
The leading cutoff effect in physical quantities 
is $O(a)$ with the Wilson quark action,
and this error can be removed by adding a single counter term, 
the Sheikholeslami-Wohlert (SW) term~\cite{SWaction},
to the action 
with non-perturbatively determined coefficient $c_{\rm SW}$.
However, $c_{\rm SW}$ has been determined only in quenched and 
two-flavor QCD so far~\cite{NPcsw.Nf0.ALPHA,NPcsw.Nf2.ALPHA}.


In this article, we perform a non-perturbative determination of 
$c_{\rm SW}$ in three-flavor QCD 
with the plaquette gauge and the Wilson quark actions.
In Refs.~\cite{PhaseDiag.Nf3.JLQCD.lat01,PhaseDiag.Nf3.JLQCD}, however,
we found that this theory has a non-trivial phase structure:
there is an unphysical phase transition 
at $\beta \! \approx \! 5.0$, 
where the lattice cutoff $a^{-1}$ is roughly 2.6~GeV.
It is expected that the phase transition
strongly distorts 
scaling properties of physical observables.
This transition is considered as an artifact due to the
finite lattice spacing and can be removed by the use of 
improved gauge actions~\cite{PhaseDiag.Nf3.JLQCD.lat01,PhaseDiag.Nf3.JLQCD}.
Therefore, there are two strategies for meaningful simulations
in three-flavor QCD:
i) use the plaquette gauge action at $\beta \gg 5.0$,
or ii) use an improved gauge action, if simulations
at $a^{-1} \lesssim 2.6$~GeV are needed.
We explore the former strategy in this article,
and the latter possibility is studied in a separate 
publication~\cite{NPcsw.Nf3.RG.CPJLQCD}.


In our determination of $c_{\rm SW}$,
we follow the method proposed by the ALPHA Collaboration
in Refs.~\cite{NPcsw.Nf0.ALPHA,NPimprovement}.
We explore a wide range of $\beta\!=\!12.0$\,--\,5.2,
which is significantly higher than the phase transition
point $\beta \! \approx \! 5.0$, 
employing a single lattice size of $8^3\times 16$.
As our main result, we derive an interpolation formula
for $c_{\rm SW}$ as a function of the bare coupling.
The critical hopping parameter $K_c$
in the $O(a)$-improved theory is also obtained
as a byproduct.
We examine finite volume effects in $c_{\rm SW}$
by modifying the improvement condition 
at one-loop order of perturbation theory.
Additional simulations in four-, two-flavor and quenched QCD
at $\beta\!=\!9.6$ are carried out 
to investigate the $N_f$ dependence of $c_{\rm SW}$.


As a preparatory step toward this study, 
we also determine $c_{\rm SW}$ in two-flavor QCD at $\beta\!=\!5.2$.
In the previous work by the ALPHA Collaboration~\cite{NPcsw.Nf2.ALPHA},
they carried out 
the non-perturbative tuning of $c_{\rm SW}$ at $\beta\!\geq\!5.4$,
and derived an interpolation formula of their result 
as a function of $g_0^2$.
However, due to the limitation of available computer power,
recent large-scale simulations 
by the UKQCD\cite{Spectrum.Nf2.UKQCD} and 
JLQCD Collaborations\cite{Spectrum.Nf2.JLQCD} 
were performed at a lower value $\beta\!=\!5.2$
with $c_{\rm SW}$ obtained by extrapolating the ALPHA's formula.
We, therefore, 
determine $c_{\rm SW}$ directly at $\beta\!=\!5.2$
in order to see if the extrapolation of the formula 
to this value of $\beta$ really works out, 
and to confirm that $O(a)$ scaling violation is absent 
in the UKQCD and JLQCD simulations.


This paper is organized as follows.
In Sec.\ref{sec:setup}, we briefly 
introduce the method for the non-pertubative tuning of $c_{\rm SW}$
employed in this study.
Section~\ref{sec:Nf2b52} is devoted to detailed description of 
our numerical analysis and results in two-flavor QCD 
at $\beta\!=\!5.2$.
We present our results in three-flavor QCD 
and discuss its $O(a/L)$ uncertainty in Sec.~\ref{sec:Nf3}.
Finally our conclusion is given in Sec.~\ref{sec:concl}.

\section{Improvement condition for $O(a)$-improvement}
\label{sec:setup}


In our determination of $c_{\rm SW}$,
we basically follow the method proposed 
in Refs.~\cite{NPcsw.Nf0.ALPHA,NPimprovement}, which employs 
the Schr\"odinger functional (SF) setup of lattice QCD~\cite{SF}.
In this section,
we briefly introduce the SF setup 
and the choice of the improvement condition to fix $c_{\rm SW}$.

\subsection{SF setup}


The SF is the generating functional of the field theory 
with the Dirichlet boundary condition
imposed in the temporal direction.
In this study, the spatial link variables at the 
boundaries are set to the following diagonal, 
constant $SU(3)$ matrices
\bea
   \left. U_k({\bf x},x_0)\right|_{x_0\!=\!0}
   = \exp \left[ aC_k \right], \ \ \ 
   \left. U_k({\bf x},x_0)\right|_{x_0\!=\!T}
   = \exp \left[ aC_k^{\prime} \right],
   \label{formul:BGF:U} \\
   C_k = \frac{i\pi}{6L_k} \left(
         \begin{array}{rrr}
            -1 & 0 & 0 \\
             0 & 0 & 0 \\
             0 & 0 & 1 
         \end{array} \right), \ \ \ 
   C_k^{\prime} = \frac{i\pi}{6L_k} \left(
         \begin{array}{rrr}
            -5 & 0 & 0 \\
             0 & 2 & 0 \\
             0 & 0 & 3 
         \end{array} \right)
   \label{formul:BGF:C},
\eea
where $L_k \ (k\!=\!1,2,3)$ and $T$ are physical 
lattice sizes in the spatial and temporal directions.
All quark fields at the boundaries are set to zero.
In the spatial directions, the periodic boundary 
condition is imposed for both gauge and quark fields.

We use the plaquette gauge action
\bea
   S_g & = & \frac{\beta}{6} 
             \sum_{x,\mu,\nu} 
             \mbox{Tr } \left[ 1 - U_{x,\mu\nu}\right],
   \label{eqn:setup:Sg}
\eea
where $U_{x,\mu\nu}$ is the product of gauge link 
variables $U_{x,\mu}$ around the plaquette 
\bea
   U_{x,\mu\nu} & = & U_{x,\mu} 
                      U_{x+\hat{\mu},\nu}
                      U_{x+\hat{\nu},\mu}^{\dagger}
                      U_{x,\nu}^{\dagger}.
   \label{eqn:setup:plaq}
\eea
The $O(a)$-improved Wilson quark action~\cite{SWaction}
is given by
\bea
   S_q & = & \sum_{x,y} \bar{q}_x D_{xy} q_y,
   \label{eqn:setup:Sq1}
   \\
   D_{xy} 
   & = & 
   \delta_{xy} 
   -K \sum_{\mu} 
    \left\{ \left( 1 - \gamma_{\mu} \right)
            U_{x,\mu} \delta_{x+\hat{\mu},y}
           +\left( 1 + \gamma_{\mu} \right)
            U_{x-\hat{\mu},\mu}^{\dagger} 
            \delta_{x-\hat{\mu},y}
                         \right\}
   +\frac{i}{2} K c_{\rm SW} \sigma_{\mu\nu} 
                             F_{x,\mu\nu} \delta_{xy},
   \label{eqn:setup:Sq2}
\eea
with the field strength tensor $F_{x,\mu\nu}$
defined by 
\bea
   F_{x,\mu\nu} 
   & = & 
   \frac{1}{8} \left\{ \left( U_{x,\mu\nu}   
                             +U_{x,\nu-\mu}
                             +U_{x,-\mu-\nu} 
                             +U_{x,-\nu\mu}
                       \right)
                      -\left(\mbox{h.c.}\right)
               \right\},
   \label{eqn:setup:Fmunu}
\eea
where $\left(\mbox{h.c.}\right)$ denotes the hermitian 
conjugate of the preceding bracket,
and $\sigma_{\mu\nu}
     =(i/2)\left[\gamma_{\mu},\gamma_{\nu}\right]$.
The last term in Eq.~(\ref{eqn:setup:Sq2}) is 
the counter term to remove $O(a)$ effects in on-shell quantities.
Its coefficient $c_{\rm SW}$ 
is set to unity to remove tree-level $O(a)$ 
scaling violation from physical observables.
The main purpose in this article is non-perturbative 
tuning of $c_{\rm SW}$ for removal of all $O(a g_0^n)$ 
scaling violation $(n \! \geq \! 0)$.
For the $O(a)$-improvement of the SF itself,
we add counter terms made of 
the gauge and quark fields at boundaries
to the lattice action.
However, these counter terms affect the PCAC relation
at order of $a^2$ or higher, and hence are not necessary 
for determination of $c_{\rm SW}$ from the PCAC relation.
In the ALPHA Collaboration's studies,
the counter terms are omitted except for 
a term
\bea
   \delta S_{g} 
   & = &
   \frac{\beta}{6} \left( c_t - 1 \right)
   \sum_{{\bf x},\mu \ (x_0=0,T\!-\!a)}
   \mbox{Tr } \left[ 1 - U_{x,\mu 0}\right],
   \label{eqn:setup:Sg:CT}
\eea
which is made of the temporal plaquettes touching the boundaries.
In this study, we also include this counter term
to the total lattice action 
$S\!=\!S_g \!+\! \delta S_g \!+\! S_q$
so that we can directly compare our and ALPHA Collaboration's results.
The coefficient of the counter term $c_t$ is set
to the one-loop estimate in Ref.~\cite{1loop_ct}.


\subsection{Improvement condition}

We determine $c_{\rm SW}$ by imposing the validity of 
the PCAC relation
\bea
   \frac{1}{2}
   \left( \partial_{\mu} + \partial_{\mu}^{*} \right)
   A_{{\rm imp},\mu}^a = 2 m P^a,
   \label{eq:setup:PCAC}
\eea
up to order of $a^2$.
The pseudo-scalar operator and $O(a)$-improved and 
unimproved axial currents are given by
\bea
   P^a 
   & = &
   \bar{\psi} \gamma_5 \tau^a \psi,
   \label{eq:setup:P}
   \\
   A_{{\rm imp},\mu}^a 
   & = &
   A_{\mu}^a + c_A \frac{1}{2} \left( \partial_{\mu} 
                                     +\partial_{\mu}^{*}
                               \right)
               P^a,
   \label{eq:setup:Aimp} 
   \\
   A_{\mu}^a 
   & = & 
   \bar{\psi} \gamma_{\mu} \gamma_5 \tau^a \psi,
   \label{eq:setup:A} 
\eea     
where $\partial_{\mu}$ and $\partial_{\mu}^{*}$ are 
the forward and backward lattice derivatives and 
$SU(N_f)$ generators $\tau^a$ 
act on the flavor indices of the quark fields
$\bar{\psi}$ and $\psi$.

We measure two correlation functions
\bea
   f_A(x_0) 
   & = &
   -\frac{1}{N_f^2-1} \langle A_0^a(x) \calO^a \rangle, 
   \label{eq:setup:fA}
   \\
   f_P(x_0) 
   & = &
   -\frac{1}{N_f^2-1} \langle P^a(x) \calO^a \rangle,
   \label{eq:setup:fP}
\eea
where $\langle\cdots\rangle$ denotes the expectation value 
after taking trace over color and spinor indices 
and summing over spatial coordinate $\bf x$.
For the source operator, we take 
\bea
   \calO^a
   & = &
   a^6 
   \sum_{\bf y, z} \bar{\zeta}({\bf y}) \gamma_5 
                   \tau^a \zeta({\bf z}),
   \label{eq:setup:source1}
\eea
defined from the boundary fields
\bea
   \zeta({\bf x}) 
   & = & \frac{\delta}{\delta \bar{\rho}({\bf x})}, \ \ \ 
   \bar{\zeta}({\bf x}) 
   = \frac{\delta}{\delta \rho({\bf x})},
   \label{eq:setup:zeta}
\eea
where $\rho({\bf x})$ is the quark field at
$x_0\!=\!0$ and is set to zero in the calculation of 
$f_A$ and $f_P$.
The bare quark mass is then calculated from $f_A$ and $f_P$
through the PCAC relation Eq.~(\ref{eq:setup:PCAC}):
\bea
   m(x_0) & = & r(x_0) + c_A s(x_0)
   \label{eq:setup:mq}
   \\
   r(x_0) & = & \frac{1}{4}
                \left( \partial_0 + \partial_0^* \right)
                f_A(x_0) / f_P(x_0)
   \label{eq:setup:r}
   \\
   s(x_0) & = & \frac{1}{2} a \, 
                \partial_0 \partial_0^*
                f_P(x_0) / f_P(x_0).
   \label{eq:setup:s}
\eea
We can calculate another set of $m^{\prime}$, $r^{\prime}$ 
and $s^{\prime}$ from the correlation functions
\bea
   f_A^{\prime}(T-x_0) 
   & = &
   +\frac{1}{N_f^2-1} \langle A_0^a(x) \calO^{\prime,a} \rangle
   \label{eq:setup:fAp},
   \\ 
   f_P^{\prime}(T-x_0) 
   & = &
   -\frac{1}{N_f^2-1} \langle P^a(x) \calO^{\prime,a} \rangle,
   \label{eq:setup:fPp}
\eea
using the source operator at the other boundary
\bea
   \calO^{\prime, a}
   & = &
   a^6 
   \sum_{\bf y, z} \bar{\zeta}^{\prime}({\bf y}) \gamma_5
                   \tau^a \zeta^{\prime}({\bf z}),
   \label{eq:setup:source2}
\eea
where $\zeta^{\prime}$  is the boundary field 
at $x_0\!=\!T$.

The improvement condition to fix $c_{\rm SW}$ is obtained by
requiring that quark masses calculated with different boundary
conditions coincide with each other.
However, a naive condition $m(x_0)\!=\!m^{\prime}(x_0)$ 
requires a non-perturbative tuning of $c_A$ as well as $c_{\rm SW}$.
To eliminate $c_A$ from the process,
it was proposed in Ref.~\cite{NPcsw.Nf0.ALPHA} 
to use a modified definition of the quark mass
\bea
   M(x_0,y_0) 
   & = &
   m(x_0) 
 - \frac{m(y_0)-m^{\prime}(y_0)}{s(y_0)-s^{\prime}(y_0)}
   s(x_0),
   \label{eq:setup:M}
\eea
and similarly defined $M^{\prime}(x_0,y_0)$.
Therefore, $c_{\rm SW}$ is tuned so that the following mass difference 
\bea 
   \Delta M(x_0,y_0) = M(x_0,y_0) - M^{\prime}(x_0,y_0)
\eea
vanishes with a certain choice of $(x_0,y_0)$.

In principle, we can take an arbitrary choice for $(x_0,y_0)$,
since a change of the choice leads to a difference
in $O(a^2)$ scaling violation in physical observables.
In this study, we take $(x_0,y_0)\!=\!(3T/4,T/4)$
for $\Delta M(x_0,y_0)$, and  $(x_0,y_0)\!=\!(T/2,T/4)$
for $M(x_0,y_0)$.
The latter is used to specify the massless point.
We note that this choice is the same 
as that in the ALPHA's studies in quenched and two-flavor QCD.
%
%
From now on, $M$ and $\Delta M$ without arguments
denote $M(T/2,T/4)$ and $\Delta M(3T/4,T/4)$, respectively.

In practice, $c_{\rm SW}$ is determined by demanding that 
$M$ and $\Delta M$ satisfy the following improvement condition
\bea 
   \left\{
   \begin{array}{lll}
      M        & = & 0, \\
      \Delta M & = & \Delta M^{(0)},
   \end{array}
   \right.
   \label{eq:setup:ImpCnd}
\eea 
where $\Delta M^{(0)}$ is the tree-level value of $\Delta M$
at the massless point $M\!=\!0$
on the finite lattice volume $L^3 \times T$.
We tune $\Delta M$ to $\Delta M^{(0)}$ but not to zero
so that the weak coupling limit of 
the non-perturbatively determined $c_{\rm SW}$ 
is exactly unity.
On our lattice size of $8^3 \times 16$, 
$a\Delta M^{(0)}\!=\!0.000277$\cite{NPcsw.Nf0.ALPHA}.
We also note that the tuning of $M$ to the massless point
provides a non-perturbative estimate 
of the critical hopping parameter $K_c$ 
in the $O(a)$-improved theory.

\section{two-flavor QCD at $\beta\!=\!5.2$}
\label{sec:Nf2b52}

\subsection{Simulation method}
\label{subsec:Nf2b52:param}

In this section, we report the determination of $c_{\rm SW}$
in two-flavor QCD at $\beta\!=\!5.2$.


Our numerical simulations are carried out
on a $8^3 \times 16$ lattice at six values of $c_{\rm SW}$ 
in a range $c_{\rm SW}\!=\!1.5$\,--\,3.0.
We choose two to four values for the hopping parameter $K$ 
at each $c_{\rm SW}$ 
so that we have data of $\Delta M$
at both of positive and negative values of $M$, 
and/or at $M$ close to the massless point 
($|aM|\!\leq\!0.01$ in our study).
This enables us to tune $(M,\Delta M)$ to $(0,\Delta M^{(0)})$
by an interpolation or short extrapolation.
The simulated values of $c_{\rm SW}$ and $K$ are 
summarized in Table~\ref{tab:Nf2b52:param}.

We use the standard HMC algorithm 
with the asymmetric even-odd preconditioning 
described in Refs.~\cite{Even-Odd0,Even-Odd,PHMC.JLQCD} 
for the determinant of the quark matrix $D$.
We solve the linear equation
$D X = B$ using the BiCGStab algorithm~\cite{BiCGStab}
with the stopping condition 
\bea
   ||R_i||/||B|| < 10^{-14},
   \label{eq:Nf2:StpCnd}
\eea
where $R_i=DX_i-B$ is the residual vector and 
$X_i$ is the estimate for the solution $X$
in the $i$-th BiCGStab iteration.
The HMC trajectory length is fixed to the unit length.
We set the number of the Molecular Dynamics steps to 60\,--\,80,
which achieves the acceptance rate higher than 80\%.

After the thermalization of 500 HMC trajectories,
we accumulate the statistics $N_{\rm traj}$ summarized in 
Table~\ref{tab:Nf2b52:param}.
The correlators $f_X$ and $f_X^{\prime} \ (X\!=\!A,P)$ are 
measured at every trajectory.
We use the jackknife method to estimate statistical errors
of $f_X$, $f_X^{\prime}$ and all results derived from them.

\subsection{numerical result}
\label{subsec:Nf2b52:result}

\begin{figure}[htbp]
   \begin{center}
   \includegraphics[width=80mm,clip]{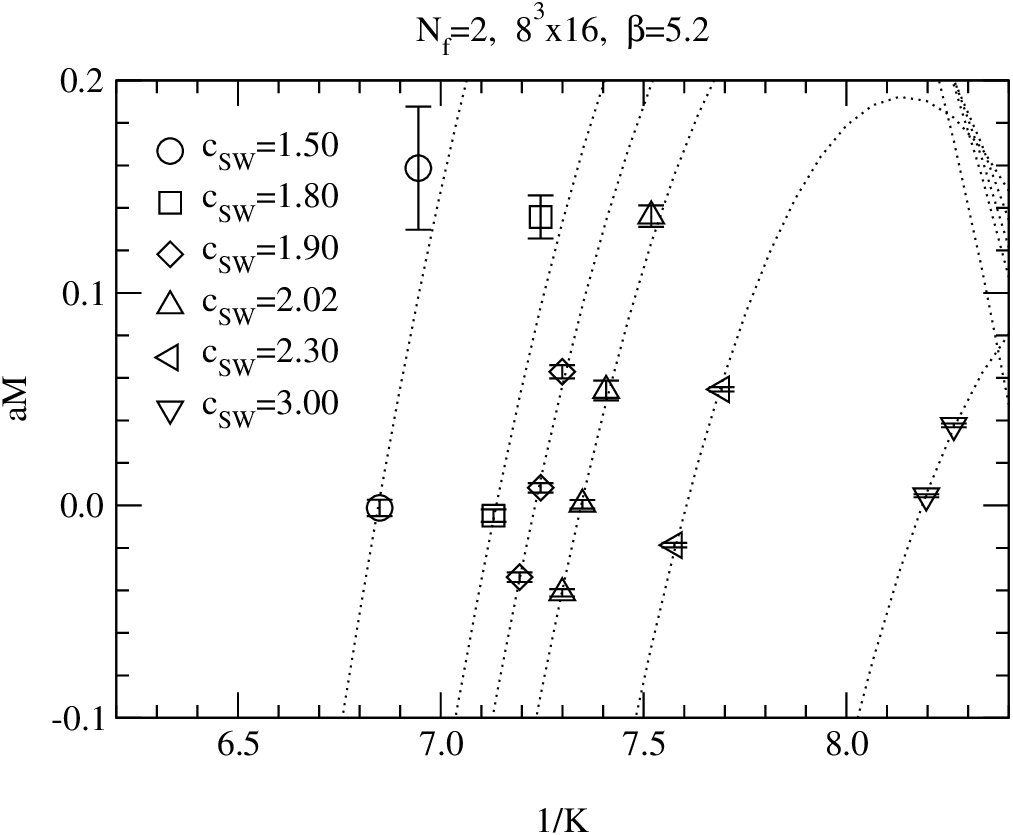}
   \end{center}
   \vspace{5mm}
   \begin{center}
   \includegraphics[width=80mm,clip]{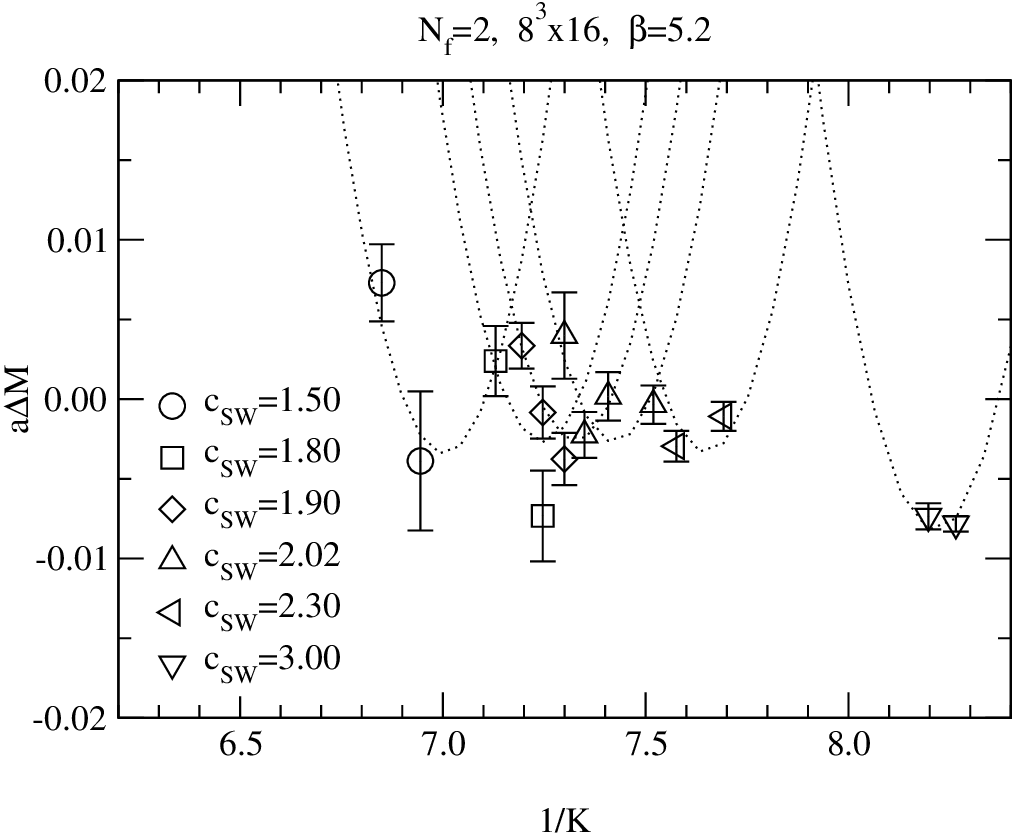}
   \end{center}
   \vspace{5mm}
   \begin{center}
   \includegraphics[width=80mm,clip]{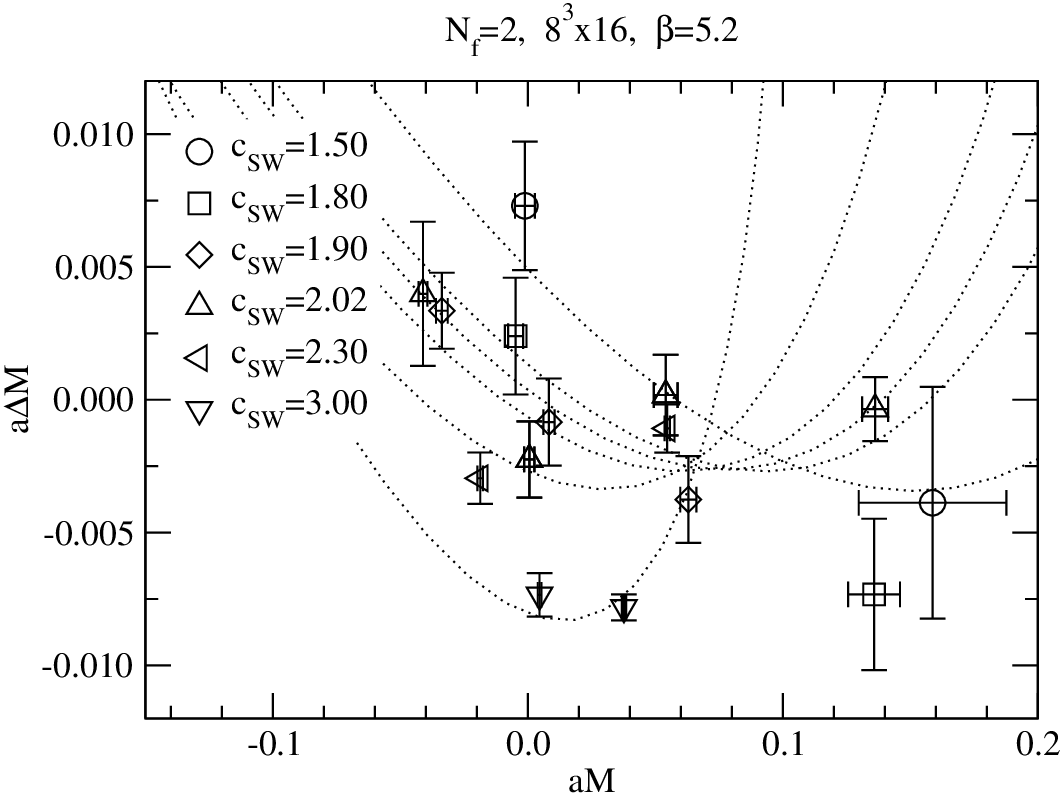}
   \end{center}
   \caption 
   {
      Plots of $M$ (top figure) and $\Delta M$ (bottom figure)
      in two-flavor QCD at $\beta\!=\!5.2$ as a function of $1/K$.
      The bottom figure shows $\Delta M$ as a function of $M$.
   }
   \label{fig:Nf2b52:MdM_vs_Kinv}
\end{figure}

Numerical results of $M$ and $\Delta M$ are summarized in
Table~\ref{tab:Nf2b52:MdM}.
%
%
%
In order to fix $c_{\rm SW}$ and $K_c$ satisfying the improvement
condition Eq.~(\ref{eq:setup:ImpCnd}),
we parameterize $M$ and $\Delta M$ 
by a simultaneous fit in terms of 
$1/K$ and $c_{\rm SW}$:
\bea
   a M        
   & = & 
   a_M + \frac{b_{M}^{(1)}}{K} + \frac{b_{M}^{(2)}}{K^2} 
       + c_M^{(1)} \, c_{\rm SW} + c_M^{(2)} \, {c_{\rm SW}}^2
       + \frac{d_M}{K}\, c_{\rm SW},
   \label{eq:Nf2b52:combfit:M} 
   \\
   a \Delta M 
   & = & 
   a_{\Delta M} + \frac{b_{\Delta M}^{(1)}}{K} 
                + \frac{b_{\Delta M}^{(2)}}{K^2} 
                + c_{\Delta M}^{(1)} \, c_{\rm SW}
                + c_{\Delta M}^{(2)} \, {c_{\rm SW}}^2
       + \frac{d_{\Delta M}}{K}\, c_{\rm SW}.
   \label{eq:Nf2b52:combfit:dM} 
\eea
Fit parameters are summarized in Table~\ref{tab:Nf2b52:MdMfit}.
Figure~\ref{fig:Nf2b52:MdM_vs_Kinv} shows
$1/K$ dependence of $M$ and $\Delta M$,
and $M$ dependence of $\Delta M$ at each $c_{\rm SW}$.
By interpolating $(M,\Delta M)$ to $(0,\Delta M^{(0)})$
with this parameterization,
we obtain 
\bea
   c_{\rm SW} & = & 1.908(64),
   \label{eq:Nf2b52:combfit:csw}
   \\
   K_c        & = & 0.1381(12).
   \label{eq:Nf2b52:combfit:Kc}
\eea

\begin{figure}[htbp]
   \begin{center}
   \includegraphics[width=80mm,clip]{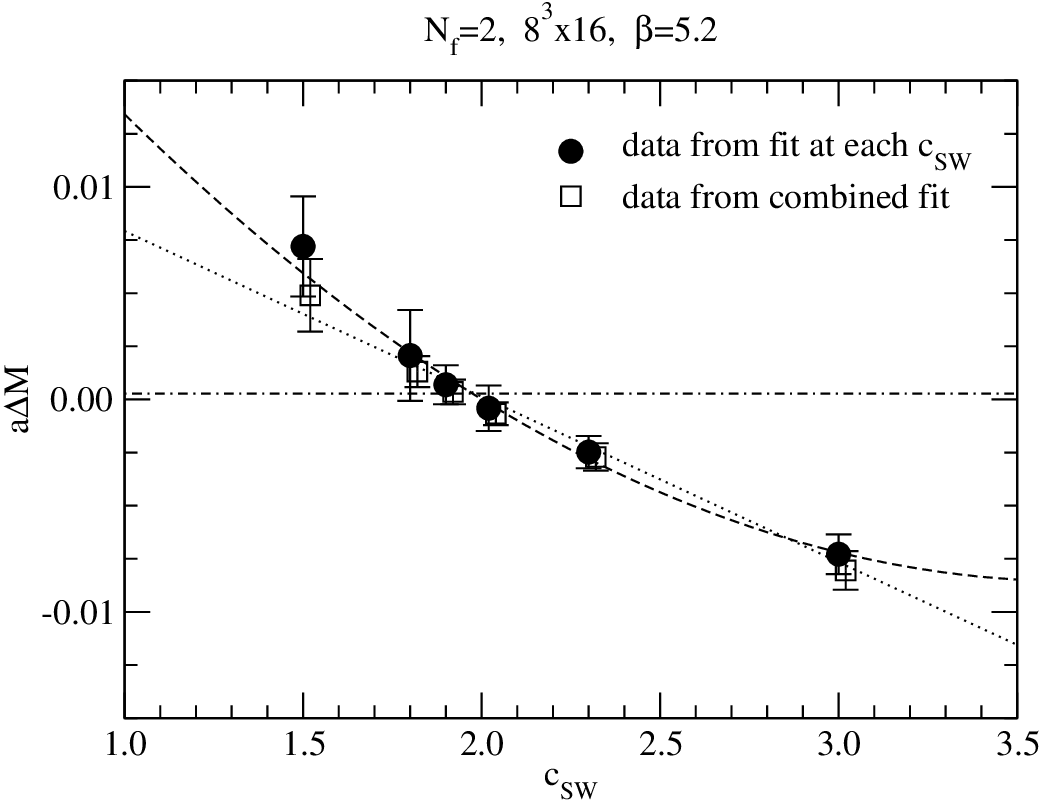}
   \end{center}
   \caption 
   {
      Mass difference $\Delta M$ at $M\!=\!0$ 
      as a function of $c_{\rm SW}$.
      Filled and open symbols are obtained from the fit
      Eq.~(\ref{eq:Nf2b52:sepafit:M})
      and combined fit
      Eqs.~(\ref{eq:Nf2b52:combfit:M}) and (\ref{eq:Nf2b52:combfit:dM}).
      Results from the combined fit are slightly shifted 
      in the horizontal direction for a better visibility.
      Dotted and dashed lines show linear and quadratic fit 
      Eq.~(\ref{eq:Nf2b52:sepafit:dM}).
      Dot-dashed line shows the tree-level value 
      $\Delta M^{(0)}\!=\!0.000277$.
   }
   \label{fig:Nf2b52:dM_vs_csw}
\end{figure}

We also test another method
for the parametrization of $M$ and $\Delta M$
in order to estimate the systematic error due to the simultaneous fit
Eqs.~(\ref{eq:Nf2b52:combfit:M}) and (\ref{eq:Nf2b52:combfit:dM}).
At each $c_{\rm SW}$, we determine $\Delta M$ at $M\!=\!0$
by a linear fit 
\bea
   a \Delta M = a_{M}^{\prime} + b_{M}^{\prime} (a M).
   \label{eq:Nf2b52:sepafit:M}
\eea
At $c_{\rm SW}\!=\!1.90$ and 2.02 where we simulate more than
two values for $K$,
we also test a quadratic form and find that 
the higher order contribution is small and can be safely neglected
in this analysis.
Figure~\ref{fig:Nf2b52:dM_vs_csw} shows 
$c_{\rm SW}$ dependence of $\Delta M$ at $M\!=\!0$
which we parametrize by linear or quadratic forms 
\bea
   a \Delta M = a_{\Delta M}^{\prime} 
              + b_{\Delta M}^{\prime}  \, c_{\rm SW}
              + c_{\Delta M}^{\prime}  \, {c_{\rm SW}}^2.
   \label{eq:Nf2b52:sepafit:dM}
\eea
By tuning $\Delta M$ to its tree-level value,
we obtain $c_{\rm SW}\!=\!1.979(68)$ from the linear fit
as reported in Ref.~\cite{Spectrum.Nf2.JLQCD},
and 1.975(50) from the quadratic one.
These are consistent with the result from the combined fit.
This good agreement 
originates from our careful choice of $c_{\rm SW}$ and $K$
in simulations:
since we choose these parameters
so that the region of $(M,\Delta M)$ 
contains $(0,\Delta M^{(0)})$
as shown in Fig.~\ref{fig:Nf2b52:MdM_vs_Kinv},
$c_{\rm SW}$ can be fixed by a short interpolation
for which the uncertainty due to the
choice of the parameterization function for $M$ and $\Delta M$
is not large.
%

The ALPHA's interpolation formula in Ref.~\cite{NPcsw.Nf2.ALPHA}
gives $c_{\rm SW}\!=\!2.017$ at $\beta\!=\!5.2$, 
which is consistent with our results.
This confirms that the ALPHA's formula can be used 
down to $\beta\!=\!5.2$
as in the UKQCD\cite{Spectrum.Nf2.UKQCD} 
and JLQCD simulations\cite{Spectrum.Nf2.JLQCD}.

However, as pointed out in Ref.~\cite{large-a}, 
there are large cutoff effects in the PCAC quark mass $m$
and the mass dependence of the Sommer scale $r_0$~\cite{r0}
around $\beta\!=\!5.2$.
%
%
%
%
%
There is a possibility that
the improvement condition Eq.~(\ref{eq:setup:ImpCnd})
adopted in this and ALPHA's previous studies 
leaves unexpectedly large $O(a^2)$ scaling violations
in physical observables 
around this value of $\beta$.
Therefore, a test of alternative improvement conditions and 
scaling properties of physical observables 
is an important subject 
to avoid the large cutoff effects in future lattice calculations.

\section{three-flavor QCD}
\label{sec:Nf3}

\subsection{Simulation method}
\label{subsec:Nf3:param}


We determine $c_{\rm SW}$ in three-flavor QCD
at nine values of $\beta$ in the range $\beta\!=\!12.0$\,--\,5.2.
Numerical simulations are carried out 
on a $8^3 \times 16$ lattice
at four values of $c_{\rm SW}$ at each $\beta$,
and three or four values of $K$ at each $c_{\rm SW}$.
These values are carefully chosen so that the region of 
$(M,\Delta M)$ contains or is sufficiently close to the point 
$(0,\Delta M^{(0)})$
which satisfies the improvement condition 
Eq.~(\ref{eq:setup:ImpCnd}).
These simulation parameters are summarized 
in Table~\ref{tab:Nf3:param}.


In our simulations,
we adopt the standard HMC algorithm for two-flavors of dynamical 
quarks 
and a polynomial HMC algorithm developed in Ref.~\cite{PHMC.JLQCD} 
for the remaining one-flavor.
We employ the symmetric even-odd preconditioning 
in Refs.~\cite{Even-Odd,PHMC.JLQCD} 
for the quark matrix $D$.
As in the two-flavor simulations at $\beta\!=\!5.2$,
we calculate $D^{-1}$
using the BiCGStab algorithm
with the tolerance parameter $||R_i||/||B|| \! < \! 10^{-14}$.
We set the number of the Molecular Dynamics steps to 80.
This achieves the acceptance rate of about 90\% or higher.

In the PHMC algorithm,
we use the Chebyshev polynomial $P[D]$ to approximate $D^{-1}$.
In order to make this algorithm exact,
the correction factor
\bea
   P_{\rm corr} & = & {\rm det}[W[D]]
   \label{eq:Nf3:PHMC:CorrFactor}
\eea
with $W[D]\!=\!P[D]D$ is taken into account
by the noisy Metropolis method~\cite{NoisyMetroP}.
We calculate the square root of $W[D]$, 
which is required in the Metropolis test, 
with an accuracy of $10^{-14}$ 
using the Taylor expansion of $W[D]$~\cite{PHMC.JLQCD}.
The order of the polynomial $N_{\rm poly}$ is chosen 
so that we achieve the acceptance rate of 
about $90$~\% or higher for the Metropolis test.

We note that, even with the SF setup,
there is a difficulty in simulating massless or negative quark masses
in three-flavor QCD.
In the strong coupling region,
eigenvalues of $D$ have large fluctuations and 
they can take values outside the radius of convergence of $P[D]$.
If this happens, the Polynomial approximation $P[D]$
and Taylor expansion of $W[D]$ break down.
For this reason, our simulations in the strong coupling region 
are performed only down to $M \! \simeq \! 0$, 
while negative quark masses $M \! \simeq \! -0.03$ are 
simulated in the weak coupling region $\beta \! \simeq \! 12$.


We accumulate statistics $N_{\rm traj}$ 
summarized in Table~\ref{tab:Nf3:param},
and measure the correlators $f_X$ and $f_X^{\prime} \ (X\!=\!A,P)$
at every trajectory.
The dependence of the jackknife error of $M$ 
on the bin size $N_{\rm bin}$ 
is investigated 
in a range $N_{\rm bin}\!=\!1$\,--\,$N_{\rm traj}/20$.
We then adopt $N_{\rm bin}$ giving the maximum error 
in the jackknife procedure in the following analysis.


We determine $c_{\rm SW}$ and $K_c$ non-perturbatively
also in four-flavor, two-flavor and quenched QCD at $\beta\!=\!9.6$ 
to study their $N_f$ dependence.
The simulation method is similar to that in three-flavor QCD,
except that we use the standard HMC algorithm in these cases.
Simulation parameters are summarized in 
Tables~\ref{tab:Nf4:param}\,--\,\ref{tab:Nf0:param}.

\subsection{Non-perturbative $c_{\rm SW}$ in three-flavor QCD}
\label{subsec:Nf3:cSW}

%

\begin{figure}[htbp]
   \begin{center}
   \includegraphics[width=80mm,clip]{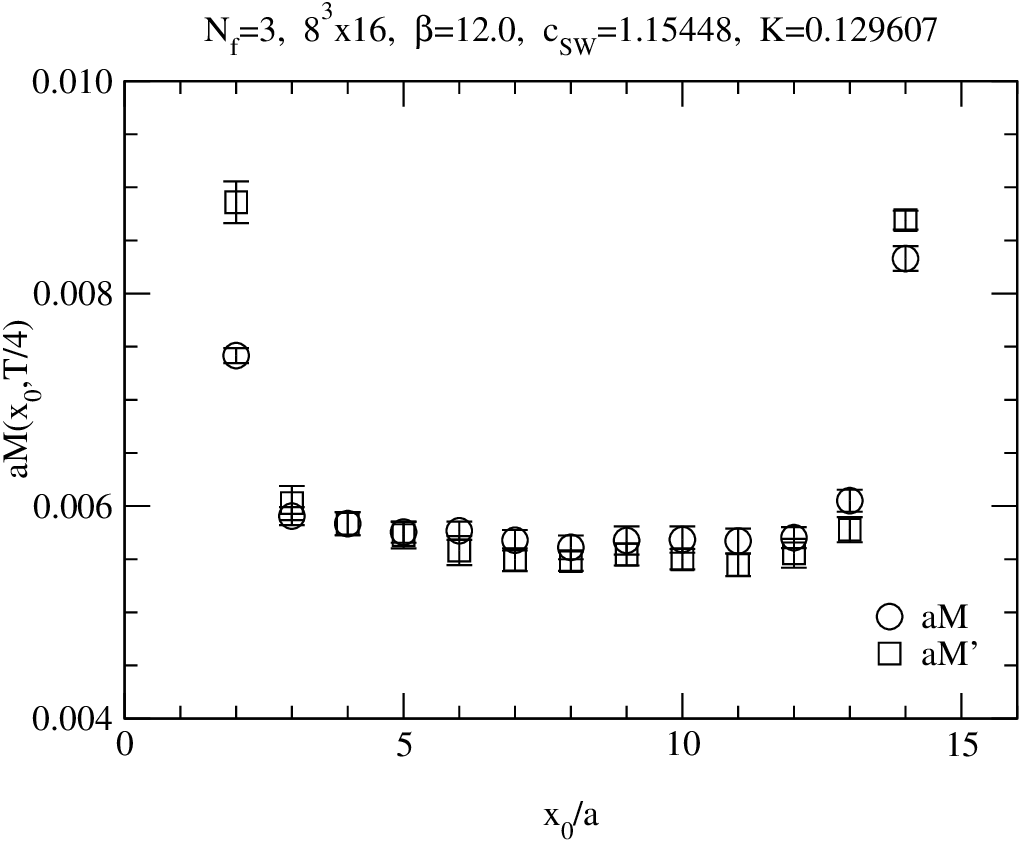}
   \end{center}
   \vspace{5mm}
   \begin{center}
   \includegraphics[width=80mm,clip]{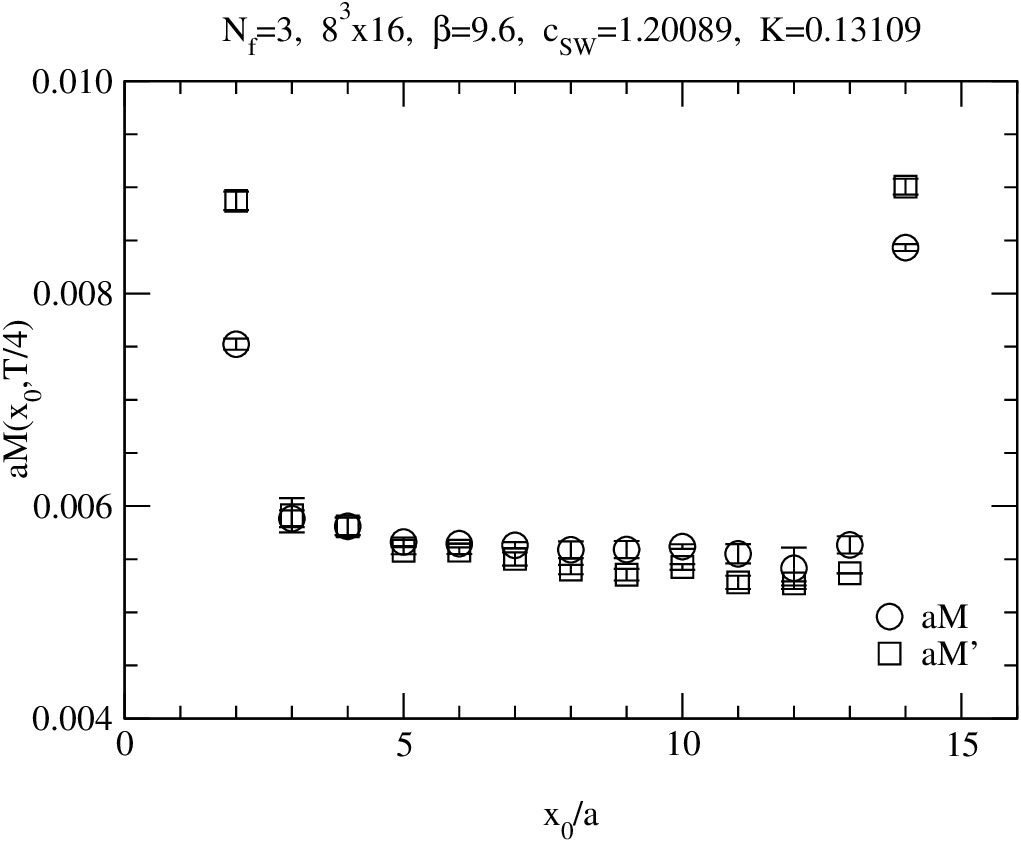}
   \end{center}
   \vspace{5mm}
   \begin{center}
   \includegraphics[width=80mm,clip]{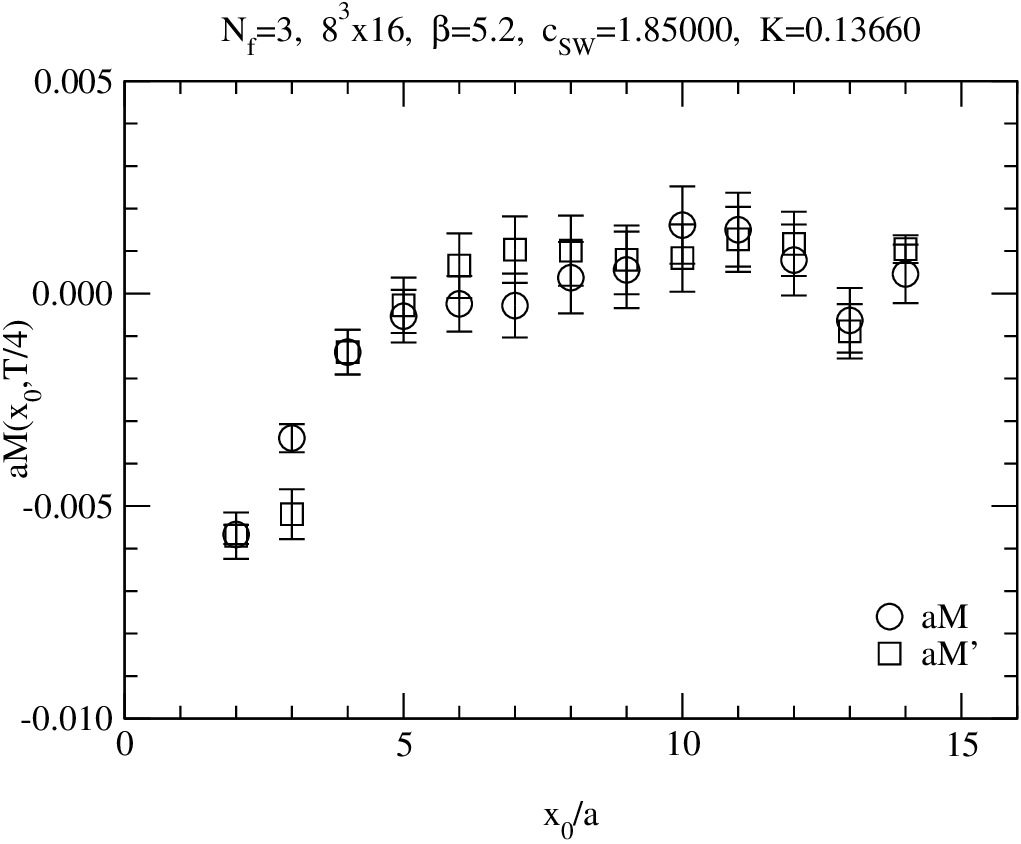}
   \end{center}
   \caption 
   {
      Quark masses $M$ and $M^{\prime}$ in three-flavor QCD
      as a function of $x_0$.
      Top, middle and bottom figures show data at $\beta\!=\!12.0$, 
      9.6 and 5.2, respectively.
   }
   \label{fig:Nf3:M_vs_x0}
\end{figure}

Numerical results of $M$ and $\Delta M$ are summarized 
in Table~\ref{tab:Nf3:MdM}.
In Fig.~\ref{fig:Nf3:M_vs_x0}, 
we plot $M$ and $M^{\prime}$ 
at several values of $\beta$ as a function of $x_0$.
With our statistics,
$aM$, $aM^{\prime}$ and hence $a\Delta M$ have
an accuracy of $10^{-3}$ at all simulation parameters.
These accurate data enables us to reduce the statistical error 
of $c_{\rm SW}$ to the level of $\sim$\,5\,\% 
even at our coarsest lattice spacing.

\begin{figure}[htbp]
   \begin{center}
   \includegraphics[width=80mm,clip]{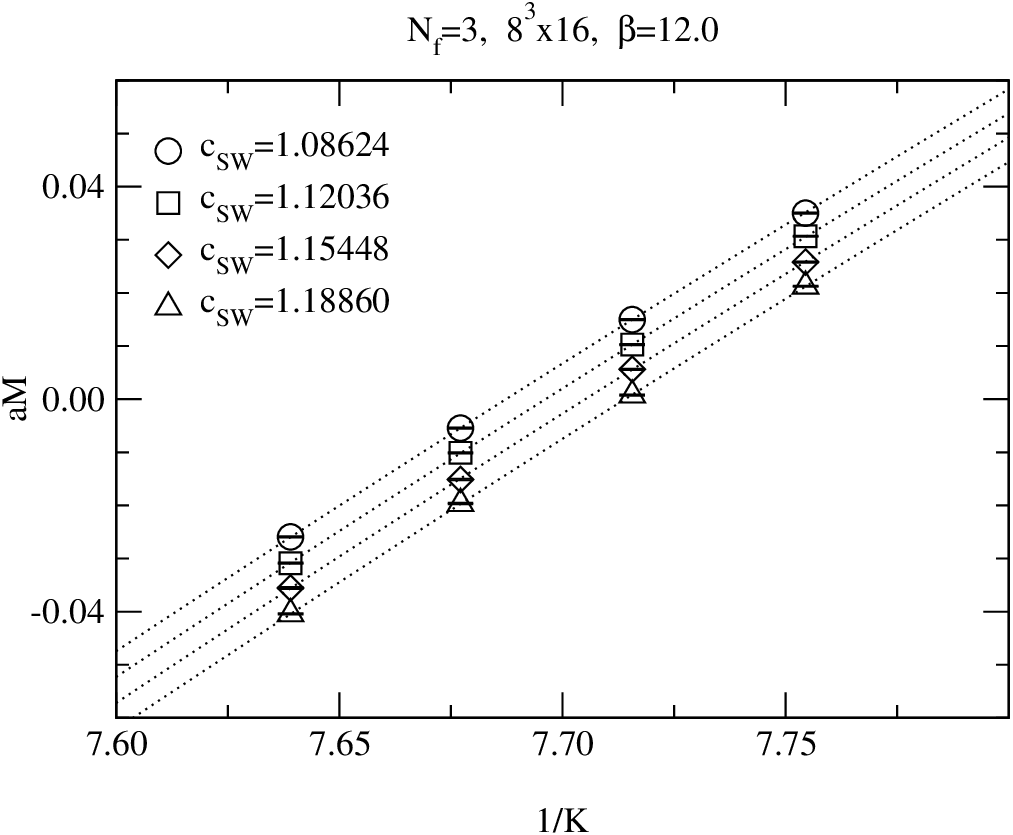}
   \end{center}
   \vspace{5mm}
   \begin{center}
   \includegraphics[width=80mm,clip]{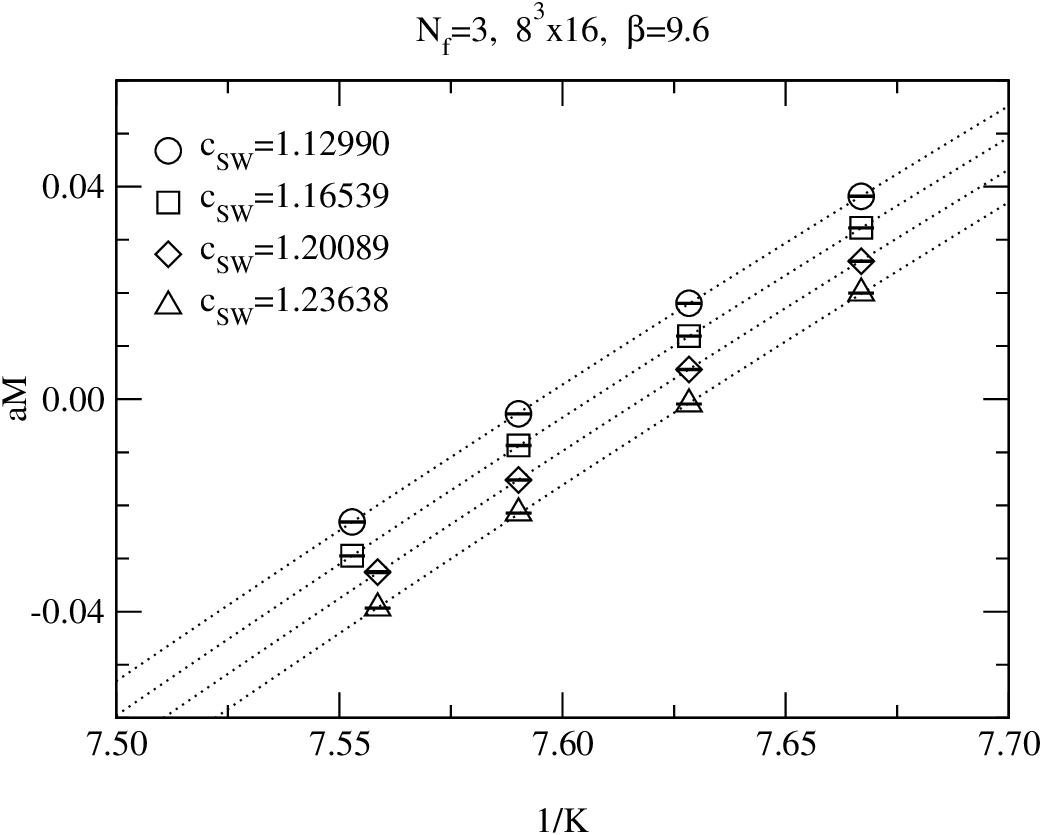}
   \end{center}
   \vspace{5mm}
   \begin{center}
   \includegraphics[width=80mm,clip]{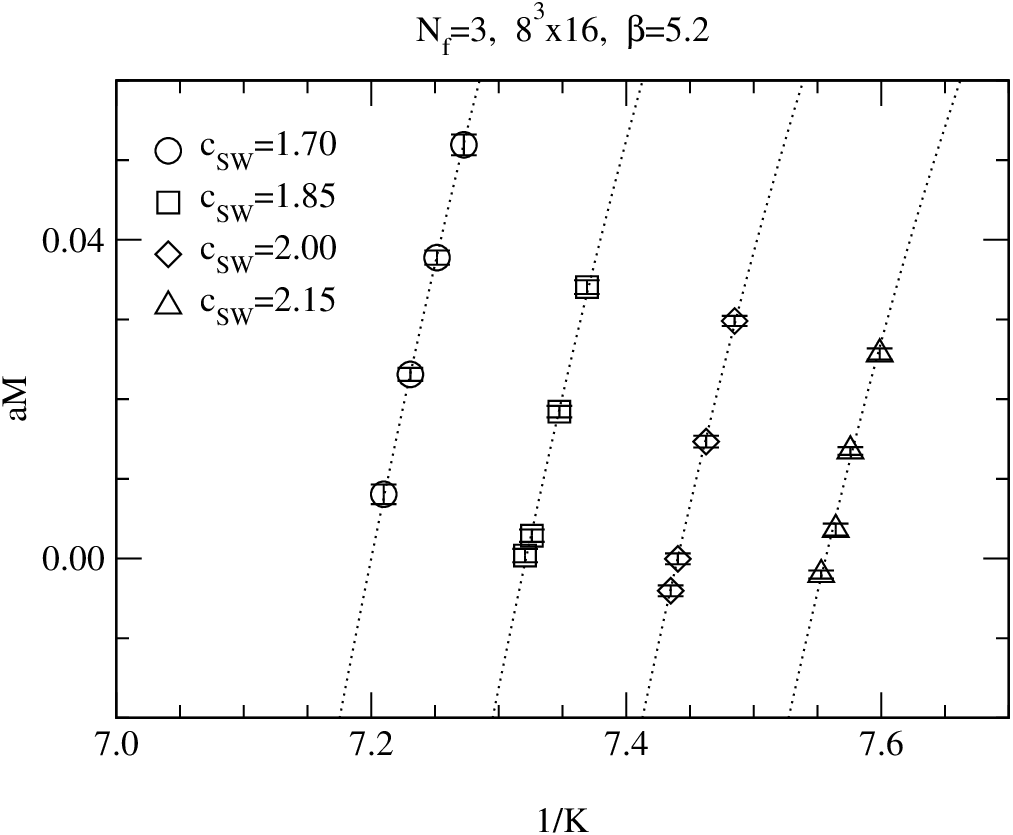}
   \end{center}
   \caption 
   {
      Quark mass $M$ as a function of $1/K$.
      Top, middle and bottom figures show data at $\beta\!=\!12.0$, 
      9.6 and 5.2, respectively.
      Dotted lines are fit curve of Eq.~(\ref{eq:Nf2b52:combfit:M}).
   }
   \label{fig:Nf3:M}
\end{figure}
\begin{figure}[htbp]
   \begin{center}
   \includegraphics[width=80mm,clip]{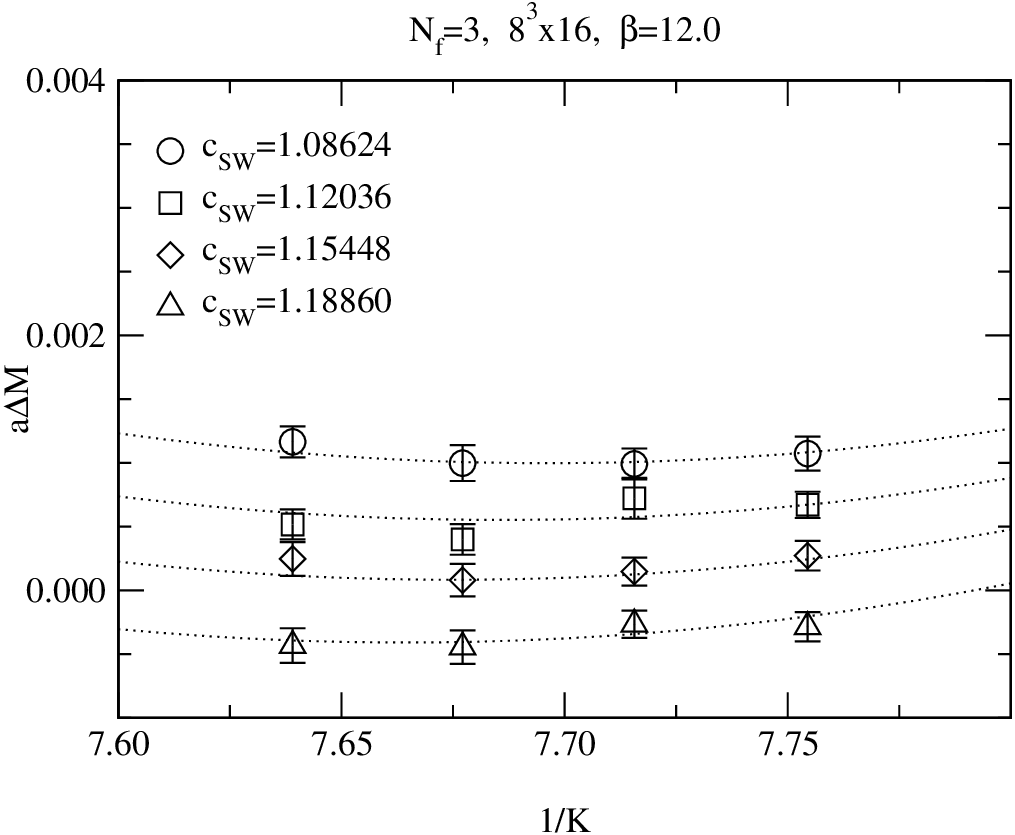}
   \end{center}
   \vspace{5mm}
   \begin{center}
   \includegraphics[width=80mm,clip]{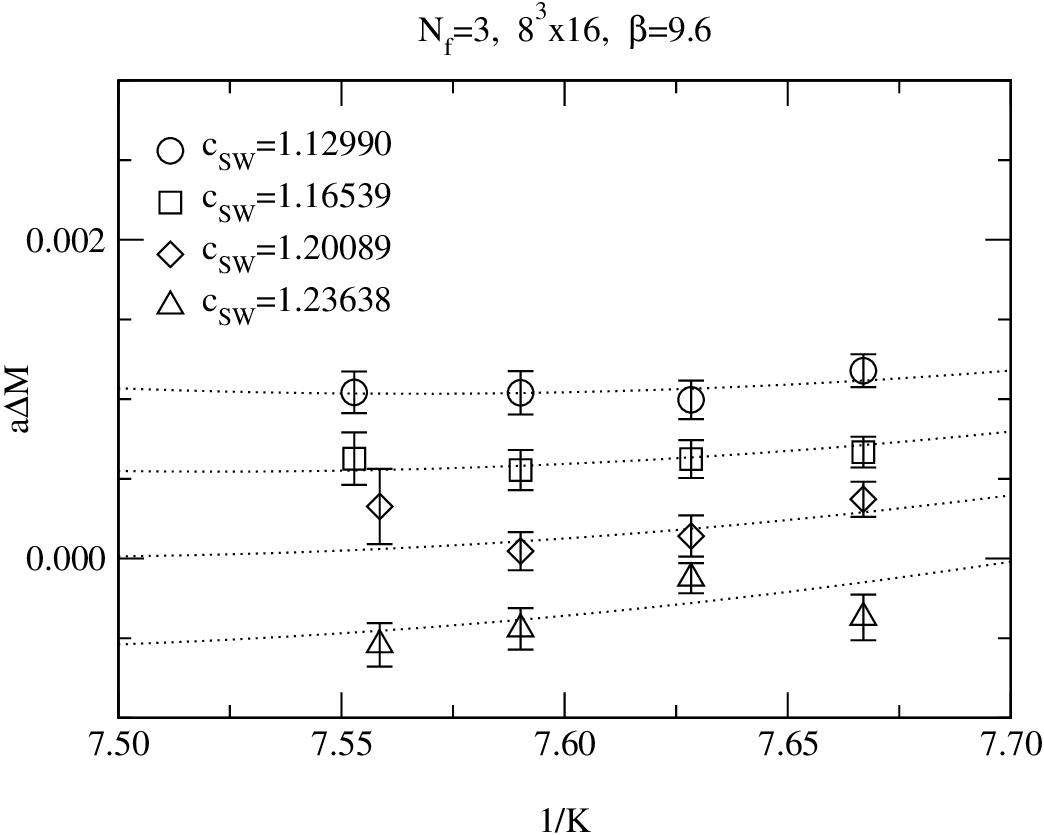}
   \end{center}
   \vspace{5mm}
   \begin{center}
   \includegraphics[width=80mm,clip]{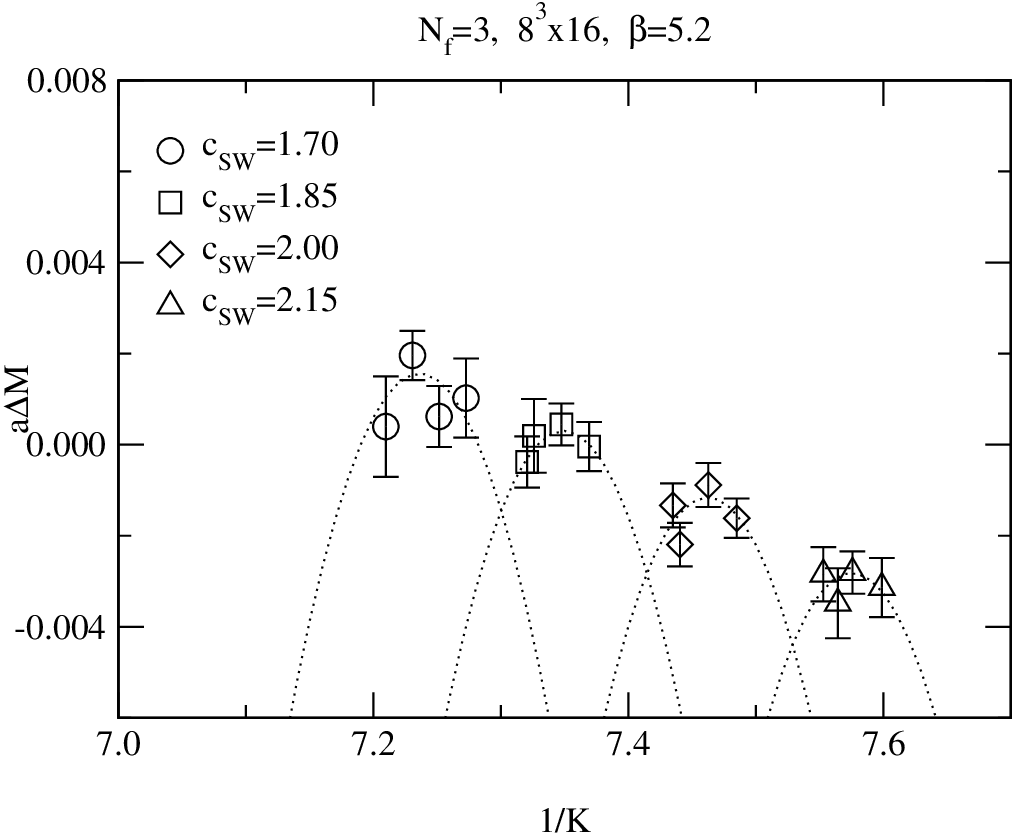}
   \end{center}
   \caption 
   {
      Mass difference $\Delta M$ as a function of $1/K$.
      Top, middle and bottom figures show data at $\beta\!=\!12.0$, 
      9.6 and 5.2, respectively.
      Dotted lines are fit curve of Eq.~(\ref{eq:Nf2b52:combfit:dM}).
   }
   \label{fig:Nf3:dM}
\end{figure}
\begin{figure}[htbp]
   \begin{center}
   \includegraphics[width=80mm,clip]{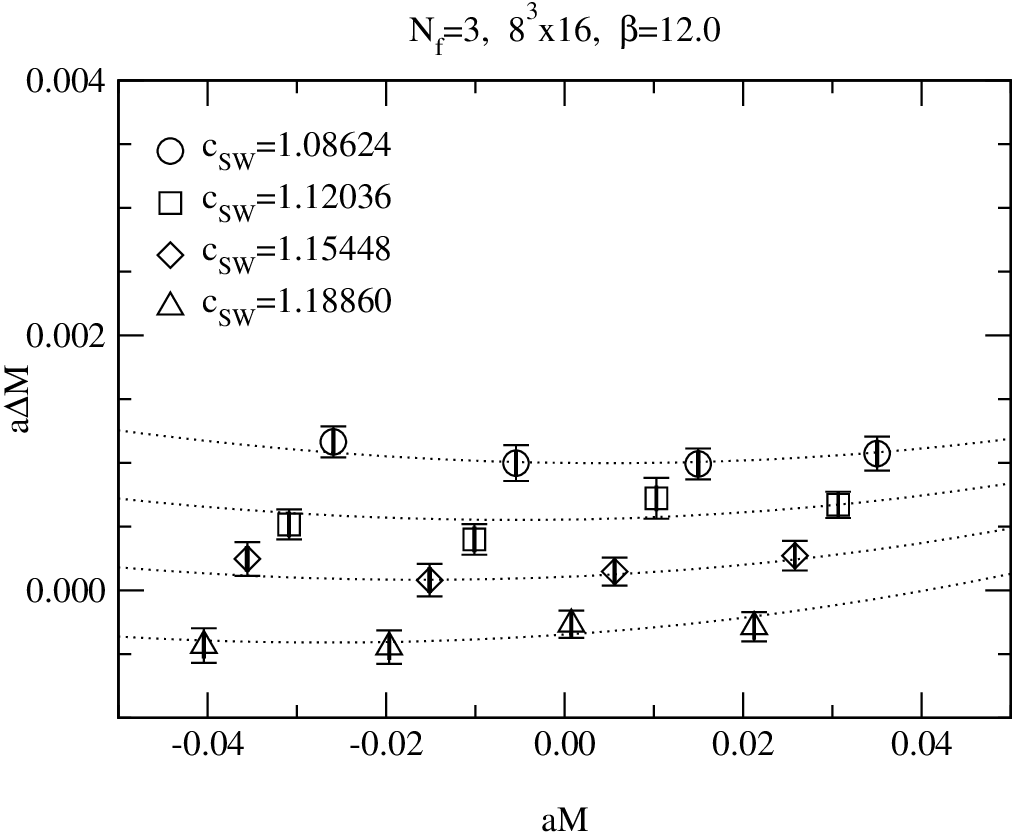}
   \end{center}
   \vspace{5mm}
   \begin{center}
   \includegraphics[width=80mm,clip]{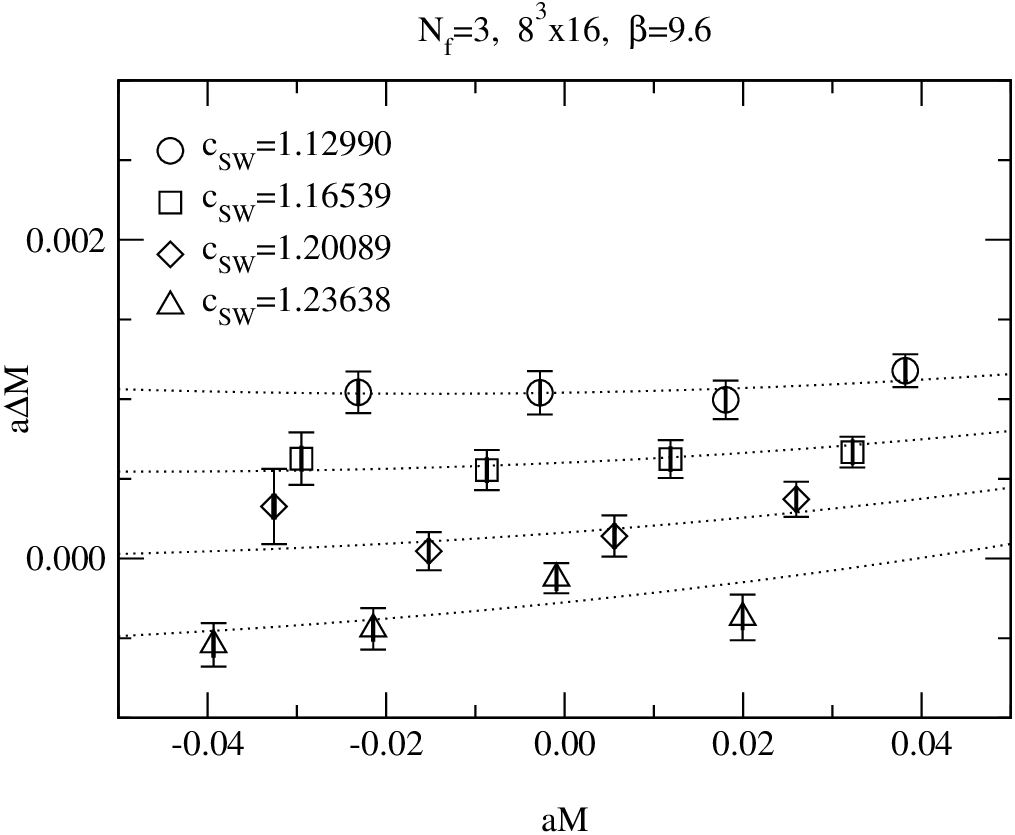}
   \end{center}
   \vspace{5mm}
   \begin{center}
   \includegraphics[width=80mm,clip]{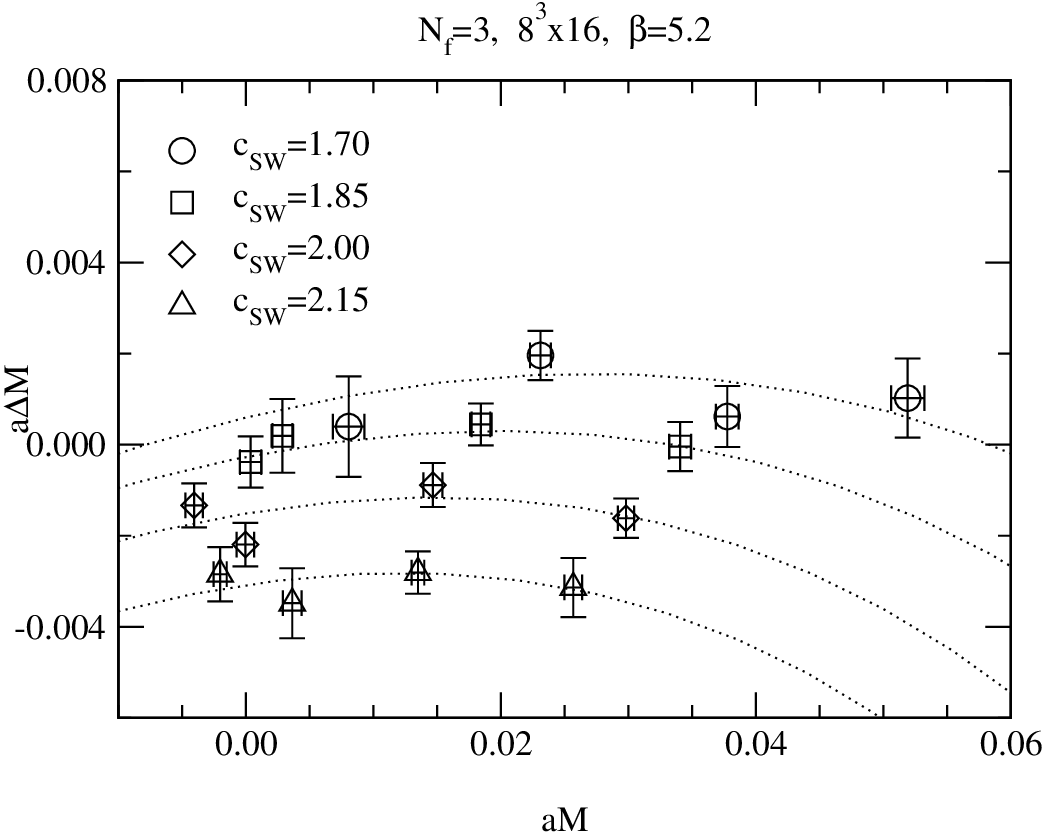}
   \end{center}
   \caption 
   {
      Mass difference $\Delta M$ as a function of $M$ 
      in three-flavor QCD.
      Top, middle and bottom figures
      show data at $\beta\!=\!12.0$, 9.6 and 5.2, respectively.
      Dotted lines are fit curve reproduced
      from Eqs.~(\ref{eq:Nf2b52:combfit:M}) and (\ref{eq:Nf2b52:combfit:dM}).
   }
   \label{fig:Nf3:dM_vs_M}
\end{figure}

In order to parameterize the $K$ and $c_{\rm SW}$ dependence of 
$M$ and $\Delta M$,
we use the combined fit of Eqs.~(\ref{eq:Nf2b52:combfit:M}) 
and (\ref{eq:Nf2b52:combfit:dM}).
Fit parameters are summarized in 
Tables~\ref{tab:Nf3:Mfit} and \ref{tab:Nf3:dMfit}.
Figures~\ref{fig:Nf3:M} and \ref{fig:Nf3:dM} show
$M$ and $\Delta M$ and their fit curves at several values of $\beta$.
We observe that our data are well described 
by the combined fit.
Consequently, 
as shown in Fig.~\ref{fig:Nf3:dM_vs_M},
$M$ dependence of $\Delta M$ is reproduced 
reasonably well by the fit.
We fix $c_{\rm SW}$ and $K_c$ satisfying the improvement 
condition Eq.~(\ref{eq:setup:ImpCnd})
using this parameterization.
Numerical results for the non-perturbatively tuned $c_{\rm SW}$ 
and $K_c$ are summarized in Table~\ref{tab:Nf3:cSW}. 

As in the analysis in two-flavor QCD at $\beta\!=\!5.2$,
we test Eqs.~(\ref{eq:Nf2b52:sepafit:M}) and 
(\ref{eq:Nf2b52:sepafit:dM}) as the alternative method 
for the parametrization.
We confirm that the two methods give consistent results 
both for $c_{\rm SW}$ and $K_c$,
and hence conclude that 
the systematic error due to the parameterization method 
for $M$ and $\Delta M$ is small.


We fit $c_{\rm SW}$ to a rational function of $g_0^2$
and obtain the following interpolation formula
\bea
   c_{\rm SW} 
   & = &
   \frac{1 - 0.194785  g_0^2 - 0.110781 g_0^4 
           - 0.0230239 g_0^6 + 0.137401 g_0^8 }
        {1 - 0.460685  g_0^2 }.
   \label{eq:Nf3:cSW} 
\eea
The interpolation formula for $K_c$ is obtained in 
a polynomial form
\bea
   K_{c} 
   & = &
   1/8 + k^{(1)} g_0^2 
       + 0.000964911 g_0^4 + 0.00298136 g_0^6 
       + 0.00100995  g_0^8 - 0.00235564 g_0^{10},
   \label{eq:Nf3:Kc} 
\eea
with $k^{(1)}\!=\!0.00843986$.
These fits reproduce our data reasonably well
with ${\chi}^2/{\rm dof}$ of around 1.6.
We also note that the coefficients of $O(g_0^2)$ terms 
in these fits are constrained so that these expressions coincide 
with their one-loop estimates~\cite{1loop_csw.W,1loop_csw.LW} 
up to $O(g_0^2)$.

\begin{figure}[htbp]
   \begin{center}
   \includegraphics[width=80mm,clip]{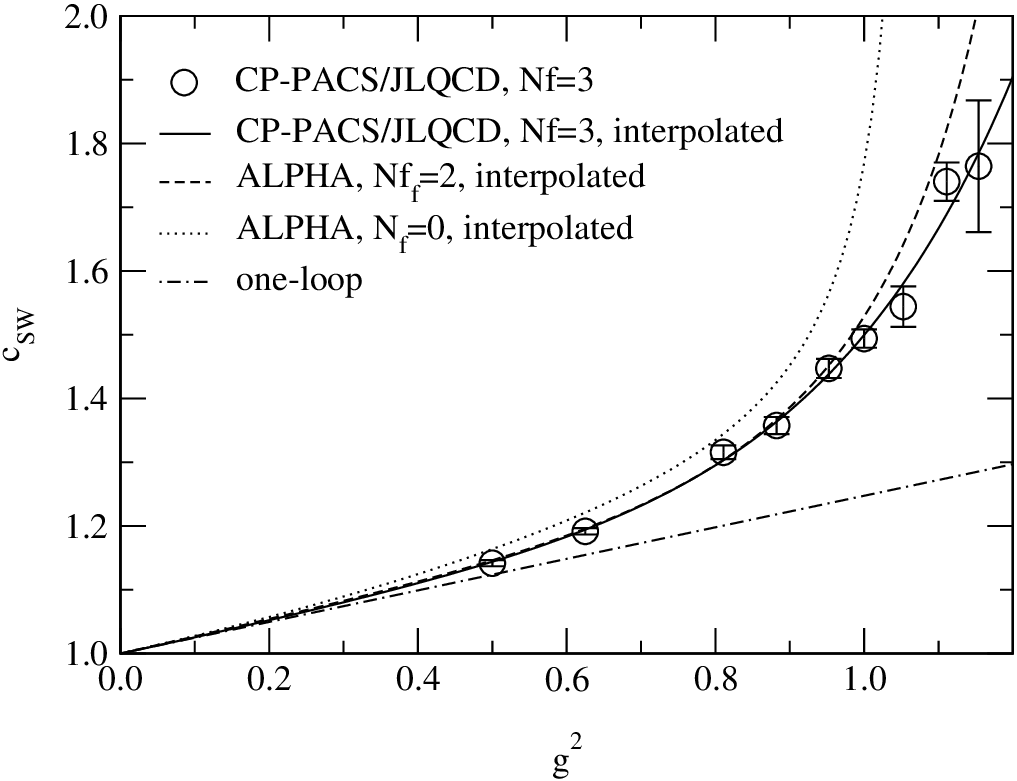}
   \end{center}
   \begin{center}
   \includegraphics[width=80mm,clip]{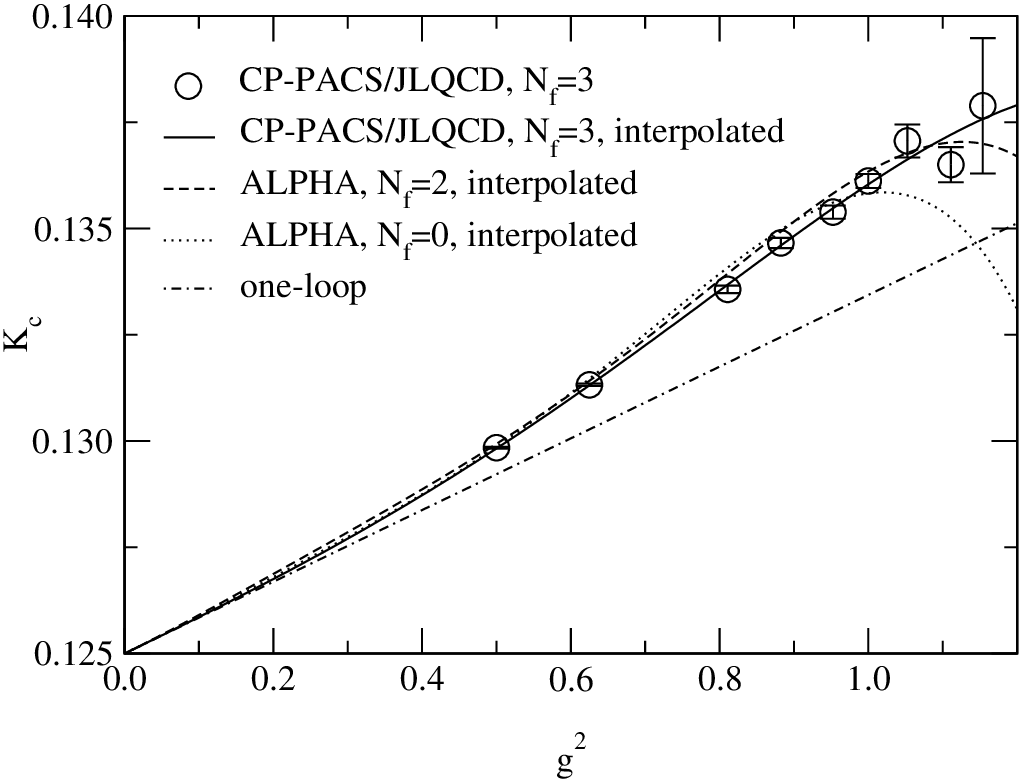}
   \end{center}
   \caption 
   {
      Non-perturbatively determined $c_{\rm SW}$ (top figure)
      and $K_c$ (bottom figure) as a function of $g_0^2$.
      For $K_c$ in quenched QCD,
      author's interpolation of the ALPHA's results is plotted.
   }
   \label{fig:Nf3:cSW_vs_g2}
\end{figure}

We plot the $\beta$ dependence of $c_{\rm SW}$ and $K_c$ 
in Fig.~\ref{fig:Nf3:cSW_vs_g2}.
While $c_{\rm SW}$ in three-flavor QCD 
is well approximated by the one-loop estimate 
in the weak coupling region of $g_0^2\! < \!0.4$ ($\beta\! > \!15$), 
it develops a significant deviation toward the strong coupling.
Similar deviation is also observed in $K_c$.
It is possible that these deviations are partly compensated
by a better choice of the expansion parameter
for lattice perturbation theory~\cite{tadpole_improvement}.
However, the large deviation at the strong coupling 
$g_0^2 \lesssim 1.0$ 
suggests that one-loop $O(a)$-improved Wilson quark 
action leads to a significant $O(a)$ scaling violation 
in physical observables at the strong coupling region,
where high statistics simulations are feasible with
currently available computer resources.
Therefore, 
the use of $c_{\rm SW}$ in Eq.~(\ref{eq:Nf3:cSW})
is essential to remove the $O(a)$ effects 
in practical lattice calculations.

\subsection{$N_f$ dependence}
\label{subsec:Nf3:Nfdep}

In Fig.~\ref{fig:Nf3:cSW_vs_g2}, 
a comparison of our interpolation formula and 
those by the ALPHA Collaboration in two-flavor and quenched QCD 
suggests that $c_{\rm SW}$ 
monotonously decreases as $N_f$ increases at fixed $\beta$.  
However, the difference between two and three-flavor QCD is not large.
This comparison may also suffer from the systematic error
due to the difference in the analysis method between 
the two collaborations.
In order to study the $N_f$ dependence more carefully,
we determine $c_{\rm SW}$ and $K_c$ for $N_f\!=\!4$, 2 and 0
at $\beta\!=\!9.6$ with an analysis method similar to that for 
$N_f\!=\!3$.
We obtain $M$ and $\Delta M$ summarized in 
Tables~\ref{tab:Nf4:MdM}\,--\,\ref{tab:Nf0:MdM}.
Fit to Eqs.~(\ref{eq:Nf2b52:combfit:M}) 
and (\ref{eq:Nf2b52:combfit:dM})
results in parameters given in Table~\ref{tab:Nf024:Mfit} 
and \ref{tab:Nf024:dMfit}.
Using the improvement condition Eq.~(\ref{eq:setup:ImpCnd}),
we obtain $c_{\rm SW}$ and $K_c$ 
summarized in Table~\ref{tab:Nf024:cSW}.

\begin{figure}[htbp]
   \begin{center}
   \includegraphics[width=80mm,clip]{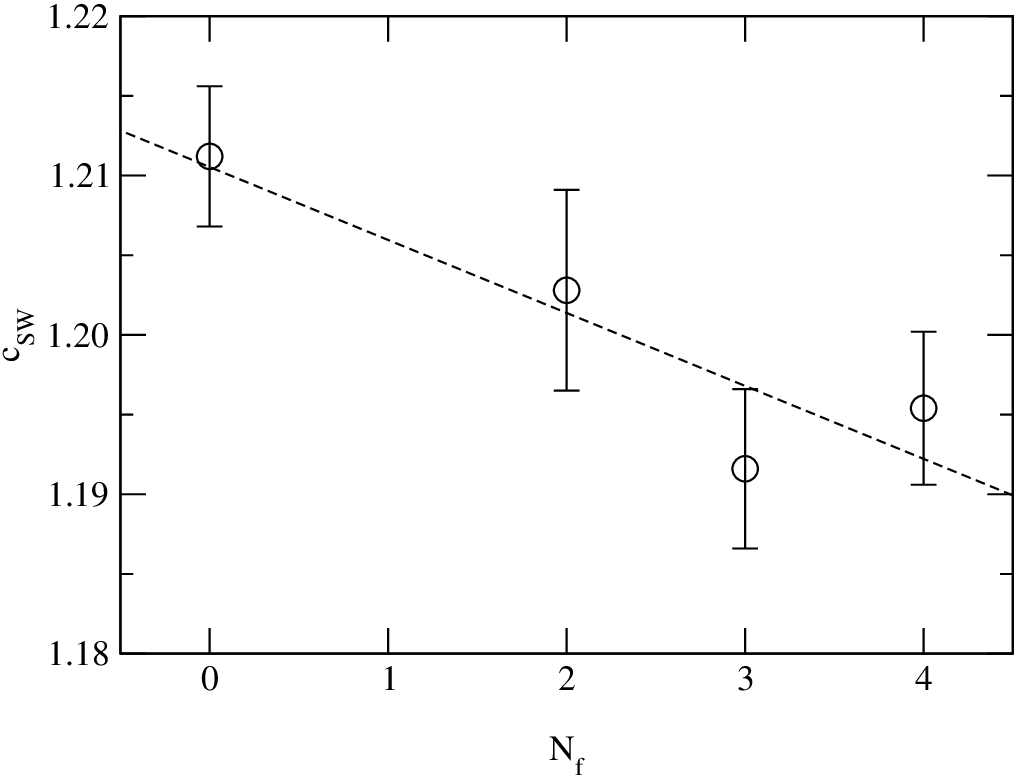}
   \end{center}
   \begin{center}
   \includegraphics[width=80mm,clip]{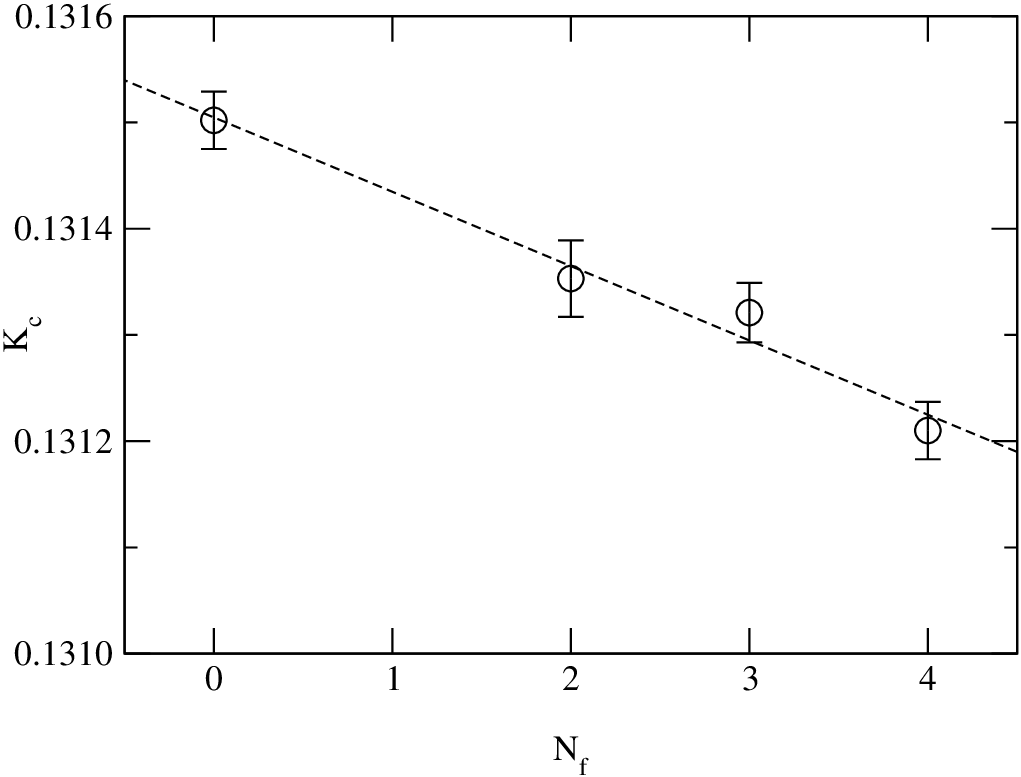}
   \end{center}
   \caption 
   {
      Plots of $c_{\rm SW}$ (top figure), $K_c$ (bottom figure)
      at $\beta\!=\!9.6$ as a function of $N_f$. 
      Dashed line shows linear fit to data.
   }
   \label{fig:Nf024:vs_Nf}
\end{figure}

Figure~\ref{fig:Nf024:vs_Nf} shows $c_{\rm SW}$ and $K_c$
at $\beta\!=\!9.6$ as a function of $N_f$.
While $K_c$ 
has an evident $N_f$ dependence,
that for $c_{\rm SW}$ is not so clear.
The leading $N_f$ dependence of $c_{\rm SW}$ is of order $g_0^4 N_f$.
By fitting our results to a linear form in $N_f$
\bea
   c_{\rm SW} & = & c_0 + c_1 g_0^4 N_f,
   \label{eq:Nf024:Nfdep}
\eea
we obtain $c_1 \!=\! -0.0117(40)$ which suggests
that the two-loop $N_f$ dependence is significant also in $c_{\rm SW}$
at the relatively weak coupling $\beta\!=\!9.6$.
If two-loop perturbative calculation of $c_{\rm SW}$ becomes available 
in future, 
it is interesting to compare the above estimate of $c_1$ 
to the perturbative estimate.


\subsection{$O(a/L)$ uncertainty in $c_{\rm SW}$}
\label{subsec:Nf3:FSE}

Our non-perturbative estimate of $c_{\rm SW}$ has 
$O(a/L)$ uncertainties, which affects physical observables 
at $O(a^2)$ when the spatial lattice size $L$ is fixed.
However, since we calculate $c_{\rm SW}$ 
with the spatial size in lattice units $L/a$ 
fixed to a constant value 8, 
the $O(a/L)$ dependence of $c_{\rm SW}$ induces 
$O(a)$ effects in observables.
These effects can be removed by extrapolating $c_{\rm SW}$
to the infinite volume limit, or interpolating to a fixed
physical size in the whole region of $g_0$. 
However, 
we are not able to do that in the present work,
since our data are taken at a single lattice size at each $g_0$.

We estimate the magnitude of $O(a/L)$ uncertainty 
in $c_{\rm SW}$ by using a modified improvement condition.
In our study as well as in ALPHA's,
the improvement condition Eq.~(\ref{eq:setup:ImpCnd}) 
is adopted instead of $\Delta M \! = \! 0$
in order to remove the tree-level $O(a/L)$ correction to $c_{\rm SW}$
\cite{NPcsw.Nf0.ALPHA}.
%
We extend this procedure to one-loop level,
namely, 
the one-loop correction to $\Delta M$
for the finite lattice volume $8^3 \times 16$ given by \cite{1loop_dM}
\bea
   a \Delta M^{(1)} 
   & = &
   - (0.00004839 + 0.00006455 N_f)
   \label{eq:OaL:dM_FSC} 
\eea
is incorporated to the improvement condition
\bea 
   \left\{
   \begin{array}{lll}
      M        & = & 0, \\
      \Delta M & = & \Delta M^{(0)} + g_0^2 \Delta M^{(1)}
   \end{array}
   \right.
   \label{eq:OaL:ImpCnd2}
\eea 
in order to remove the $O(g_0^2 a/L)$ correction from $c_{\rm SW}$.

From the parametrization of
Eqs.~(\ref{eq:Nf2b52:combfit:M})\,--\,(\ref{eq:Nf2b52:combfit:dM})
and the modified improvement condition Eq.~(\ref{eq:OaL:ImpCnd2}),
we obtain $c_{\rm SW}$ and $K_c$ summarized 
in Table~\ref{tab:FSC:cSW}.
A comparison with Table~\ref{tab:Nf3:cSW} shows
that the modified and original improvement conditions 
give consistent results for $c_{\rm SW}$ 
with each other,
and hence that 
the $O(g_0^2 a/L)$ correction is small in our results.

In Fig.~\ref{fig:Nf3:M_vs_x0}, 
we observe that $M$ and $M^{\prime}$ 
have a mild $x_0$ dependence at $4a < x_0 < 14a$.
This may suggest that 
different choices of $x_0$ for the improvement conditions
Eqs.~(\ref{eq:setup:ImpCnd}) and (\ref{eq:OaL:ImpCnd2})
lead to a small difference in $c_{\rm SW}$
and hence its $O(a/L)$ ambiguity is not large.

From these observations, we expect that 
$O(g_0^n a/L)$ corrections are not large at $L/a\!=\!8$,
and that $c_{\rm SW}$ in the infinite volume limit is well 
approximated by our results.
It is an important subject in future studies 
to confirm this point by a direct calculation 
of $c_{\rm SW}$ with varying $L$.

\section{Conclusion}
\label{sec:concl}

In this paper, we have performed a non-perturbative 
$O(a)$-improvement of the Wilson quark action
in three-flavor QCD with the plaquette gauge action.
Our high statistics at carefully chosen simulation parameters 
$c_{\rm SW}$ and $K$ enable us to determine 
non-perturbative $c_{\rm SW}$ with an accuracy of $\sim$\,5\,\% level
in the wide range of $\beta\!=\!12.0$\,--\,5.2.
The main result of this study is the interpolation formula
Eq.~(\ref{eq:Nf3:cSW}), with which 
$O(a)$ scaling violation in physical observables 
can be removed in future simulations at $\beta \! \geq \! 5.2$.
As a byproduct, we also obtain the interpolation formula for $K_c$,
which is useful to locate simulation points.

While it is expected that $O(a/L)$ uncertainty in $c_{\rm SW}$ 
is not large, this point should be confirmed 
by a direct determination of $c_{\rm SW}$ with varying $L$.
A test of scaling properties of physical observables
with our estimate of $c_{\rm SW}$ is an important next step 
toward an extensive simulation of three-flavor QCD 
with the plaquette gauge action.

We note that 
the range of $\beta$ explored in this study
is significantly higher than $\beta\!=\!5.0$, 
where an artificial phase transition exists.
Numerical studies at $\beta$ below our range 
would possibly suffer from large distortion 
of scaling properties of physical observables.
The use of improved gauge actions removes the 
lattice artifact and can push 
simulations toward coarser lattice spacings.
This possibility is explored in a separate 
publication~\cite{NPcsw.Nf3.RG.CPJLQCD}.


\begin{acknowledgments}
This work is supported by the Supercomputer Project No.98 (FY2003)
of High Energy Accelerator Research Organization (KEK),
and also in part by the Grant-in-Aid of the Ministry of Education
(Nos. 12740133, 13135204, 13640260, 14046202, 14740173, 15204015,
15540251, 15540279, 16028201, 16540228, 16740147).

\end{acknowledgments}



\clearpage


\begin{table}
\begin{center}
\caption{Simulation parameters in two-flavor QCD at $\beta$=5.2.}
\begin{ruledtabular}
\begin{tabular}{ccc||ccc||ccc}
   $c_{\rm SW}$ & $K$      & $N_{\rm traj}$  &
   $c_{\rm SW}$ & $K$      & $N_{\rm traj}$  &
   $c_{\rm SW}$ & $K$      & $N_{\rm traj}$ 
   \\ \hline
   1.50  & 0.14400  & 6400    &    1.90  & 0.13800  & 5600    &   2.02  & 0.13700  & 4800    \\
   1.50  & 0.14600  & 6200    &    1.90  & 0.13900  & 6387    &   2.30  & 0.13000  & 3200    \\
   1.80  & 0.13800  & 6400    &    2.02  & 0.13300  & 8000    &   2.30  & 0.13200  & 4000    \\
   1.80  & 0.14025  & 6400    &    2.02  & 0.13500  & 2176    &   3.00  & 0.12100  & 4000    \\
   1.90  & 0.13700  & 6400    &    2.02  & 0.13607  & 6400    &   3.00  & 0.12200  & 4000    
 \end{tabular}	     
\end{ruledtabular}	     
\label{tab:Nf2b52:param}
\vspace{-5ex}
\end{center}
\end{table}

\begin{table}
\begin{center}
\caption{Quark mass $M$ and its difference $\Delta M$ 
         in two-flavor QCD at $\beta$=5.2.}
\begin{ruledtabular}
\begin{tabular}{cccc||cccc||cccc}
   $c_{\rm SW}$ & $K$      & $aM$ & $a\Delta M$ &
   $c_{\rm SW}$ & $K$      & $aM$ & $a\Delta M$ &
   $c_{\rm SW}$ & $K$      & $aM$ & $a\Delta M$ 
   \\ \hline
   1.50  & 0.14400  &  0.159(29)    & -0.0039(44)   &
   1.90  & 0.13800  &  0.0083(22)   & -0.0008(16)   &
   2.02  & 0.13700  & -0.0411(17)   &  0.0040(27)   
   \\
   1.50  & 0.14600  & -0.0012(39)   &  0.0073(24)   &
   1.90  & 0.13900  & -0.0338(23)   &  0.0033(14)   &
   2.30  & 0.13000  &  0.0546(10)   & -0.00108(91)  
   \\   
   1.80  & 0.13800  &  0.136(10)    & -0.0073(28)   &
   2.02  & 0.13300  &  0.1361(51)   & -0.0003(12)   &
   2.30  & 0.13200  & -0.0187(11)   & -0.00296(97)  
   \\   
   1.80  & 0.14025  & -0.0048(29)   &  0.0024(22)   &
   2.02  & 0.13500  &  0.0541(47)   &  0.0002(15)   &
   3.00  & 0.12100  &  0.03767(84)  & -0.00782(49)  
   \\   
   1.90  & 0.13700  &  0.0629(31)   & -0.0038(16)   &
   2.02  & 0.13607  &  0.0005(20)   & -0.0022(14)   &
   3.00  & 0.12200  &  0.00455(71)  & -0.00735(82)  
\end{tabular}	     
\end{ruledtabular}	     
\label{tab:Nf2b52:MdM}
\vspace{-5ex}
\end{center}
\end{table}

\begin{table}
\begin{center}
\caption{
   Fit parameters for Eqs.~(\ref{eq:Nf2b52:combfit:M}) and 
   (\ref{eq:Nf2b52:combfit:dM}) in two-flavor QCD at $\beta\!=\!5.2$.
}
\begin{ruledtabular}
\begin{tabular}{llllllll}
   $\chi^2/{\rm dof}$ 
 & $a_M$          & $b_{M}^{(1)}$  & $b_{M}^{(2)}$
                  & $c_{M}^{(1)}$  & $c_{M}^{(2)}$ 
                  & $d_{M}$
\\ \hline
   6.37  & -28.8(7.1) &  8.9(2.5)  & -0.67(22) 
                      & -6.3(2.2)  & -0.15(17)
                      &  0.83(39)
\\ \hline
   $\chi^2/{\rm dof}$ 
 & $a_{\Delta M}$ & $b_{\Delta M}^{(1)}$ & $b_{\Delta M}^{(2)}$
                  & $c_{\Delta M}^{(1)}$ & $c_{\Delta M}^{(2)}$ 
                  & $d_{\Delta M}$
\\ \hline
   1.88  & 11.2(3.2)  & -3.9(1.1)  &  0.336(98)
                      &  3.16(95)  &  0.214(71)
                      & -0.54(17)
\end{tabular}
\end{ruledtabular}
\label{tab:Nf2b52:MdMfit}
\vspace{-5ex}
\end{center}
\end{table}

\clearpage

\begin{table}
\begin{center}
\caption{Simulation parameters in three-flavor QCD.}

\begin{ruledtabular}
\begin{tabular}{cc||cc||cc||cc}
   \multicolumn{8}{c}{$\beta\!=\!12.0$}
   \\ \hline
   \multicolumn{2}{c||}{$c_{\rm SW}\!=\!1.08624$} & 
   \multicolumn{2}{c||}{$c_{\rm SW}\!=\!1.12036$} & 
   \multicolumn{2}{c||}{$c_{\rm SW}\!=\!1.15448$} & 
   \multicolumn{2}{c  }{$c_{\rm SW}\!=\!1.18860$} 
   \\ \hline
   $K$      & $N_{\rm traj}$  &
   $K$      & $N_{\rm traj}$  &
   $K$      & $N_{\rm traj}$  &
   $K$      & $N_{\rm traj}$ 
   \\ \hline
   0.128958 & 3000 & 0.128958 & 3000 & 0.128958 & 3000 & 0.128958 & 3000
   \\
   0.129607 & 3000 & 0.129607 & 3000 & 0.129607 & 3000 & 0.129607 & 3000
   \\
   0.130257 & 3000 & 0.130257 & 3000 & 0.130257 & 3000 & 0.130257 & 3000
   \\
   0.130906 & 3000 & 0.130906 & 3000 & 0.130906 & 3000 & 0.130906 & 3000
%
   \\ \hline
   \multicolumn{8}{c}{$\beta\!=\!9.6$}
   \\ \hline
   \multicolumn{2}{c||}{$c_{\rm SW}\!=\!1.12990$} & 
   \multicolumn{2}{c||}{$c_{\rm SW}\!=\!1.16539$} & 
   \multicolumn{2}{c||}{$c_{\rm SW}\!=\!1.20089$} & 
   \multicolumn{2}{c  }{$c_{\rm SW}\!=\!1.23638$} 
   \\ \hline
   $K$      & $N_{\rm traj}$  &
   $K$      & $N_{\rm traj}$  &
   $K$      & $N_{\rm traj}$  &
   $K$      & $N_{\rm traj}$ 
   \\ \hline
   0.13043 & 7000 &  0.13043 & 7000 &  0.13043 & 7000 &  0.13043 & 7000 
   \\
   0.13109 & 7000 &  0.13109 & 7000 &  0.13109 & 7000 &  0.13109 & 7000 
   \\
   0.13175 & 7000 &  0.13175 & 7000 &  0.13175 & 7000 &  0.13175 & 7000 
   \\
   0.13240 & 7000 &  0.13240 & 7000 &  0.13230 & 1800 &  0.13230 & 7000 
%
   \\ \hline
   \multicolumn{8}{c}{$\beta\!=\!7.4$}
   \\ \hline
   \multicolumn{2}{c||}{$c_{\rm SW}\!=\!1.2258$} & 
   \multicolumn{2}{c||}{$c_{\rm SW}\!=\!1.2643$} & 
   \multicolumn{2}{c||}{$c_{\rm SW}\!=\!1.3028$} & 
   \multicolumn{2}{c  }{$c_{\rm SW}\!=\!1.3413$} 
   \\ \hline
   $K$      & $N_{\rm traj}$  &
   $K$      & $N_{\rm traj}$  &
   $K$      & $N_{\rm traj}$  &
   $K$      & $N_{\rm traj}$ 
   \\ \hline
   0.13293 & 5400 & 0.13293 & 5400 & 0.13293 & 5400 & 0.13293 & 5400 
   \\
   0.13360 & 5400 & 0.13360 & 5400 & 0.13360 & 5400 & 0.13360 & 5400 
   \\
   0.13427 & 5400 & 0.13427 & 5400 & 0.13427 & 5400 & 0.13393 &  400 
   \\
   0.13494 & 5400 & 0.13470 & 3400 & --      & --   & 0.13427 & 5400 
%
   \\ \hline
   \multicolumn{8}{c}{$\beta\!=\!6.8$}
   \\ \hline
   \multicolumn{2}{c||}{$c_{\rm SW}\!=\!1.2783$} & 
   \multicolumn{2}{c||}{$c_{\rm SW}\!=\!1.3184$} & 
   \multicolumn{2}{c||}{$c_{\rm SW}\!=\!1.3586$} & 
   \multicolumn{2}{c  }{$c_{\rm SW}\!=\!1.3987$} 
   \\ \hline
   $K$      & $N_{\rm traj}$  &
   $K$      & $N_{\rm traj}$  &
   $K$      & $N_{\rm traj}$  &
   $K$      & $N_{\rm traj}$ 
   \\ \hline
   0.13391 & 4200 & 0.13391 & 4200 & 0.13391 & 4200 & 0.13391 & 4200
   \\
   0.13459 & 4200 & 0.13459 & 4200 & 0.13459 & 4200 & 0.13459 & 4200
   \\
   0.13526 & 4200 & 0.13526 & 4200 & 0.13526 & 4200 & 0.13492 & 3700
   \\
   0.13594 & 4200 & 0.13560 & 4200 & 0.13540 & 3700 & 0.13500 & 3150
%
   \\ \hline
   \multicolumn{8}{c}{$\beta\!=\!6.3$}
   \\ \hline
   \multicolumn{2}{c||}{$c_{\rm SW}\!=\!1.3117$} & 
   \multicolumn{2}{c||}{$c_{\rm SW}\!=\!1.3675$} & 
   \multicolumn{2}{c||}{$c_{\rm SW}\!=\!1.4233$} & 
   \multicolumn{2}{c  }{$c_{\rm SW}\!=\!1.4791$} 
   \\ \hline
   $K$      & $N_{\rm traj}$  &
   $K$      & $N_{\rm traj}$  &
   $K$      & $N_{\rm traj}$  &
   $K$      & $N_{\rm traj}$ 
   \\ \hline
   0.13501 & 4600 & 0.13501 & 4600 & 0.13446 & 4600 & 0.13446 & 4600 
   \\
   0.13555 & 4600 & 0.13555 & 4600 & 0.13501 & 4600 & 0.13501 & 4600 
   \\
   0.13609 & 4600 & 0.13609 & 4600 & 0.13555 & 4600 & 0.13530 & 1100 
   \\
   0.13664 & 4600 & 0.13664 & 4600 & 0.13609 & 4600 & 0.13555 & 4600 
%
   \\ \hline
   \multicolumn{8}{c}{$\beta\!=\!6.0$}
   \\ \hline
   \multicolumn{2}{c||}{$c_{\rm SW}\!=\!1.3237$} & 
   \multicolumn{2}{c||}{$c_{\rm SW}\!=\!1.3801$} & 
   \multicolumn{2}{c||}{$c_{\rm SW}\!=\!1.4364$} & 
   \multicolumn{2}{c  }{$c_{\rm SW}\!=\!1.4927$} 
   \\ \hline
   $K$      & $N_{\rm traj}$  &
   $K$      & $N_{\rm traj}$  &
   $K$      & $N_{\rm traj}$  &
   $K$      & $N_{\rm traj}$ 
   \\ \hline
   0.13552 & 4200 & 0.13552 & 4200 & 0.13498 & 4200 & 0.13498 & 4200 
   \\
   0.13607 & 4200 & 0.13607 & 4200 & 0.13552 & 4200 & 0.13552 & 4200 
   \\
   0.13661 & 4200 & 0.13661 & 4200 & 0.13607 & 4200 & 0.13607 & 4200 
   \\
   0.13716 & 4200 & 0.13716 & 4200 & 0.13661 & 4200 & 0.13634 & 4200 
%
   \\ \hline
   \multicolumn{8}{c}{$\beta\!=\!5.7$}
   \\ \hline
   \multicolumn{2}{c||}{$c_{\rm SW}\!=\!1.5431$} & 
   \multicolumn{2}{c||}{$c_{\rm SW}\!=\!1.6088$} & 
   \multicolumn{2}{c||}{$c_{\rm SW}\!=\!1.6745$} & 
   \multicolumn{2}{c  }{$c_{\rm SW}\!=\!1.7401$} 
   \\ \hline
   $K$      & $N_{\rm traj}$  &
   $K$      & $N_{\rm traj}$  &
   $K$      & $N_{\rm traj}$  &
   $K$      & $N_{\rm traj}$ 
   \\ \hline
   0.13575 & 5000 & 0.13541 & 5000 & 0.13507 & 5000 & 0.13400 & 4500
   \\
   0.13644 & 5000 & 0.13610 & 5000 & 0.13540 & 4400 & 0.13450 & 4100
   \\
   0.13712 & 4700 & 0.13644 & 3700 & 0.13575 & 5000 & 0.13473 & 5000
   \\
   0.13750 & 4500 & 0.13678 & 5000 & 0.13585 & 2700 & 0.13500 & 4400
\end{tabular}	     
\end{ruledtabular}	     

\label{tab:Nf3:param}
\vspace{-5ex}
\end{center}
\end{table}

\addtocounter{table}{-1}
\begin{table}
\begin{center}
\caption{({\it Continued})}

\begin{ruledtabular}
\begin{tabular}{cc||cc||cc||cc}
   \multicolumn{8}{c}{$\beta\!=\!5.4$}
   \\ \hline
   \multicolumn{2}{c||}{$c_{\rm SW}\!=\!1.6$} & 
   \multicolumn{2}{c||}{$c_{\rm SW}\!=\!1.7$} & 
   \multicolumn{2}{c||}{$c_{\rm SW}\!=\!1.8$} & 
   \multicolumn{2}{c  }{$c_{\rm SW}\!=\!1.9$} 
   \\ \hline
   $K$      & $N_{\rm traj}$  &
   $K$      & $N_{\rm traj}$  &
   $K$      & $N_{\rm traj}$  &
   $K$      & $N_{\rm traj}$ 
   \\ \hline
   0.13750 & 9000 & 0.13600 & 9000 & 0.13480 & 9000 & 0.13330 & 9000
   \\
   0.13810 & 9000 & 0.13630 & 7900 & 0.13520 & 9000 & 0.13370 & 9000
   \\
   0.13825 & 7100 & 0.13660 & 9000 & 0.13560 & 9000 & 0.13410 & 9000
   \\
   0.13840 & 8200 & 0.13720 & 9000 & 0.13580 & 7700 & 0.13450 & 9000
%
   \\ \hline
   \multicolumn{8}{c}{$\beta\!=\!5.2$}
   \\ \hline
   \multicolumn{2}{c||}{$c_{\rm SW}\!=\!1.70$} & 
   \multicolumn{2}{c||}{$c_{\rm SW}\!=\!1.85$} & 
   \multicolumn{2}{c||}{$c_{\rm SW}\!=\!2.00$} & 
   \multicolumn{2}{c  }{$c_{\rm SW}\!=\!2.15$} 
   \\ \hline
   $K$      & $N_{\rm traj}$  &
   $K$      & $N_{\rm traj}$  &
   $K$      & $N_{\rm traj}$  &
   $K$      & $N_{\rm traj}$ 
   \\ \hline
   0.1375 & 7800 & 0.1357 & 7800 & 0.1336 & 7800 & 0.1316 & 7800 
   \\
   0.1379 & 7800 & 0.1361 & 7800 & 0.1340 & 7800 & 0.1320 & 7800 
   \\
   0.1383 & 7800 & 0.1365 & 7800 & 0.1344 & 7800 & 0.1322 & 5500
   \\
   0.1387 & 6300 & 0.1366 & 7400 & 0.1345 & 7400 & 0.1324 & 7800
\end{tabular}	     
\end{ruledtabular}	     

\label{tab:Nf3:param2}
\vspace{-5ex}
\end{center}
\end{table}

\begin{table}
\begin{center}
\caption{Simulation parameters in four-flavor QCD at $\beta\!=\!9.6$.}

\begin{ruledtabular}
\begin{tabular}{cc||cc||cc||cc}
   \multicolumn{2}{c||}{$c_{\rm SW}\!=\!1.12990$} & 
   \multicolumn{2}{c||}{$c_{\rm SW}\!=\!1.16539$} & 
   \multicolumn{2}{c||}{$c_{\rm SW}\!=\!1.20089$} & 
   \multicolumn{2}{c  }{$c_{\rm SW}\!=\!1.23638$} 
   \\ \hline
   $K$      & $N_{\rm traj}$  &
   $K$      & $N_{\rm traj}$  &
   $K$      & $N_{\rm traj}$  &
   $K$      & $N_{\rm traj}$ 
   \\ \hline
   0.13043 & 5000 & 0.13043 & 5000 & 0.13043 & 5000 & 0.13043 & 5000
   \\
   0.13109 & 5000 & 0.13109 & 5000 & 0.13109 & 5000 & 0.13109 & 5000
   \\
   0.13175 & 5000 & 0.13175 & 5000 & 0.13175 & 5000 & 0.13175 & 5000
   \\
   0.13240 & 5000 & 0.13240 & 5000 & 0.13240 & 5000 & 0.13240 & 5000
\end{tabular}	     
\end{ruledtabular}	     

\label{tab:Nf4:param}
\vspace{-5ex}
\end{center}
\end{table}

\begin{table}
\begin{center}
\caption{Simulation parameters in two-flavor QCD at $\beta\!=\!9.6$.}

\begin{ruledtabular}
\begin{tabular}{cc||cc||cc||cc}
   \multicolumn{2}{c||}{$c_{\rm SW}\!=\!1.12990$} & 
   \multicolumn{2}{c||}{$c_{\rm SW}\!=\!1.16539$} & 
   \multicolumn{2}{c||}{$c_{\rm SW}\!=\!1.20089$} & 
   \multicolumn{2}{c  }{$c_{\rm SW}\!=\!1.23638$} 
   \\ \hline
   $K$      & $N_{\rm traj}$  &
   $K$      & $N_{\rm traj}$  &
   $K$      & $N_{\rm traj}$  &
   $K$      & $N_{\rm traj}$ 
   \\ \hline
   0.13043 & 5000 & 0.13043 & 5000 & 0.13043 & 5000 & 0.13043 & 5000
   \\
   0.13109 & 5000 & 0.13109 & 5000 & 0.13109 & 5000 & 0.13109 & 5000
   \\
   0.13175 & 5000 & 0.13175 & 5000 & 0.13175 & 5000 & 0.13175 & 5000
   \\
   0.13240 & 5000 & 0.13240 & 5000 & 0.13240 & 5000 & 0.13240 & 5000
\end{tabular}	     
\end{ruledtabular}	     

\label{tab:Nf2:param}
\vspace{-5ex}
\end{center}
\end{table}

\begin{table}
\begin{center}
\caption{Simulation parameters in quenched QCD at $\beta\!=\!9.6$.}

\begin{ruledtabular}
\begin{tabular}{cc||cc||cc||cc}
   \multicolumn{2}{c||}{$c_{\rm SW}\!=\!1.12990$} & 
   \multicolumn{2}{c||}{$c_{\rm SW}\!=\!1.16539$} & 
   \multicolumn{2}{c||}{$c_{\rm SW}\!=\!1.20089$} & 
   \multicolumn{2}{c  }{$c_{\rm SW}\!=\!1.23638$} 
   \\ \hline
   $K$      & $N_{\rm traj}$  &
   $K$      & $N_{\rm traj}$  &
   $K$      & $N_{\rm traj}$  &
   $K$      & $N_{\rm traj}$ 
   \\ \hline
   0.13043 & 8000 & 0.13043 & 8000 & 0.13043 & 8000 & 0.13043 & 8000
   \\
   0.13109 & 8000 & 0.13109 & 8000 & 0.13109 & 8000 & 0.13109 & 8000
   \\
   0.13175 & 8000 & 0.13175 & 8000 & 0.13175 & 8000 & 0.13175 & 8000
   \\
   0.13240 & 8000 & 0.13240 & 8000 & 0.13240 & 8000 & 0.13240 & 8000
\end{tabular}	     
\end{ruledtabular}

\label{tab:Nf0:param}
\vspace{-5ex}
\end{center}
\end{table}


\clearpage


\begin{table}
\begin{center}
\caption{Quark mass $M$ and its difference $\Delta M$ 
         in three-flavor QCD.}

\begin{ruledtabular}
\begin{tabular}{ccc||ccc}
   \multicolumn{6}{c}{$\beta\!=\!12.0$}
   \\ \hline
   \multicolumn{3}{c||}{$c_{\rm SW}\!=\!1.08624$} & 
   \multicolumn{3}{c}{$c_{\rm SW}\!=\!1.12036$} 
   \\ \hline
   $K$      & $aM$ & $a\Delta M$ &
   $K$      & $aM$ & $a\Delta M$
   \\ \hline
   0.128958 &   0.03502(11) &   0.00107(13)   & 
   0.128958 &   0.03066(8)  &   0.00067(10)   
   \\
   0.129607 &   0.01498(13) &   0.00099(12)   & 
   0.129607 &   0.01028(10) &   0.00072(16)   
   \\
   0.130257 &$-$0.00544(10) &   0.00100(14)   & 
   0.130257 &$-$0.01011(12) &   0.00040(12)   
   \\
   0.130906 &$-$0.02591(11) &   0.00116(12)   & 
   0.130906 &$-$0.03088(11) &   0.00052(12)   
   \\ \hline
   \multicolumn{3}{c||}{$c_{\rm SW}\!=\!1.15448$} & 
   \multicolumn{3}{c  }{$c_{\rm SW}\!=\!1.18860$} 
   \\ \hline
   $K$      & $aM$ & $a\Delta M$ &
   $K$      & $aM$ & $a\Delta M$ 
   \\ \hline
   0.128958 &   0.02581(9)  &   0.00027(12)   &  
   0.128958 &   0.02125(9)  &$-$0.00029(11)   
   \\
   0.129607 &   0.00561(11) &   0.00015(11)   &  
   0.129607 &   0.00076(10) &$-$0.00027(11)   
   \\
   0.130257 &$-$0.01512(14) &   0.00008(13)   &  
   0.130257 &$-$0.01965(11) &$-$0.00044(13)   
   \\
   0.130906 &$-$0.03555(17) &   0.00025(13)   &
   0.130906 &$-$0.04041(12) &$-$0.00043(14)   
%
   \\ \hline
   \multicolumn{6}{c}{$\beta\!=\!9.6$}
   \\ \hline
   \multicolumn{3}{c||}{$c_{\rm SW}\!=\!1.12990$} & 
   \multicolumn{3}{c  }{$c_{\rm SW}\!=\!1.16539$} 
   \\ \hline
   $K$      & $aM$ & $a\Delta M$ &
   $K$      & $aM$ & $a\Delta M$
   \\ \hline
   0.13043 &   0.03818(12) &   0.00118(10) & 0.13043 &   0.03226(9)  &   0.00067(10) 
   \\
   0.13109 &   0.01805(9)  &   0.00100(12) & 0.13109 &   0.01187(10) &   0.00062(12) 
   \\
   0.13175 &$-$0.00275(10) &   0.00104(14) & 0.13175 &$-$0.00871(12) &   0.00056(13) 
   \\
   0.13240 &$-$0.02312(11) &   0.00104(13) & 0.13240 &$-$0.02948(11) &   0.00063(16) 
   \\ \hline
   \multicolumn{3}{c||}{$c_{\rm SW}\!=\!1.20089$} & 
   \multicolumn{3}{c  }{$c_{\rm SW}\!=\!1.23638$} 
   \\ \hline
   $K$      & $aM$ & $a\Delta M$ &
   $K$      & $aM$ & $a\Delta M$ 
   \\ \hline
   0.13043 &   0.02596(11) &   0.00037(11) & 0.13043 &   0.01995(9)  &$-$0.00037(14) 
   \\
   0.13109 &   0.00559(12) &   0.00014(13) & 0.13109 &$-$0.00091(10) &$-$0.00012(9)  
   \\
   0.13175 &$-$0.01522(11) &   0.00005(12) & 0.13175 &$-$0.02145 (11)&$-$0.00044(13) 
   \\
   0.13230 &$-$0.03256(19) &   0.00033(24) & 0.13230 &$-$0.03934 (13)&$-$0.00054(14) 
%
   \\ \hline
   \multicolumn{6}{c}{$\beta\!=\!7.4$}
   \\ \hline
   \multicolumn{3}{c||}{$c_{\rm SW}\!=\!1.2258$} & 
   \multicolumn{3}{c  }{$c_{\rm SW}\!=\!1.2643$} 
   \\ \hline
   $K$      & $aM$ & $a\Delta M$ &
   $K$      & $aM$ & $a\Delta M$
   \\ \hline
   0.13293 &   0.04147(22) &   0.00135(19)  & 0.13293 &   0.03265(15) &   0.00067(19) 
   \\
   0.13360 &   0.02134(21) &   0.00158(25)  & 0.13360 &   0.01209(23) &   0.00065(25) 
   \\
   0.13427 &   0.00079(21) &   0.00099(26)  & 0.13427 &$-$0.00942(21) &   0.00102(23) 
   \\
   0.13494 &$-$0.02065(24) &   0.00077(28)  & 0.13470 &$-$0.02292(28) &   0.00089(34) 
   \\ \hline
   \multicolumn{3}{c||}{$c_{\rm SW}\!=\!1.3028$} & 
   \multicolumn{3}{c  }{$c_{\rm SW}\!=\!1.3413$} 
   \\ \hline
   $K$      & $aM$ & $a\Delta M$ &
   $K$      & $aM$ & $a\Delta M$ 
   \\ \hline
   0.13293 &   0.02296(25) &   0.00035(24) & 0.13293 &   0.01309(22) &   0.00004(23) 
   \\
   0.13360 &   0.00214(17) &   0.00018(19) & 0.13360 &$-$0.00756(21) &   0.00029(28) 
   \\
   0.13427 &$-$0.01888(23) &   0.00052(22) & 0.13393 &$-$0.01774(64) &$-$0.00018(68) 
   \\
   --      &   --          &   --          & 0.13427 &$-$0.02867(26) &$-$0.00048(36) 
%
   \\ \hline
   \multicolumn{6}{c}{$\beta\!=\!6.8$}
   \\ \hline
   \multicolumn{3}{c||}{$c_{\rm SW}\!=\!1.2783$} & 
   \multicolumn{3}{c  }{$c_{\rm SW}\!=\!1.3184$} 
   \\ \hline
   $K$      & $aM$ & $a\Delta M$ &
   $K$      & $aM$ & $a\Delta M$
   \\ \hline
   0.13391 &   0.04462(36) &   0.00066(28) & 0.13391 &   0.03342(33) &   0.00092(28) 
   \\
   0.13459 &   0.02457(29) &   0.00079(32) & 0.13459 &   0.01358(28) &   0.00051(27) 
   \\
   0.13526 &   0.00338(33) &   0.00098(30) & 0.13526 &$-$0.00823(30) &   0.00065(40) 
   \\
   0.13594 &$-$0.01824(30) &   0.00163(37) & 0.13560 &$-$0.01839(36) &   0.00054(39) 
   \\ \hline
   \multicolumn{3}{c||}{$c_{\rm SW}\!=\!1.3586$} & 
   \multicolumn{3}{c  }{$c_{\rm SW}\!=\!1.3987$} 
   \\ \hline
   $K$      & $aM$ & $a\Delta M$ &
   $K$      & $aM$ & $a\Delta M$ 
   \\ \hline
   0.13391 &   0.02240(31) &   0.00049(26) & 0.13391 &   0.01208(22) &$-$0.00021(35) 
   \\
   0.13459 &   0.00175(31) &   0.00031(23) & 0.13459 &$-$0.00953(30) &$-$0.00030(27) 
   \\
   0.13526 &$-$0.01924(27) &   0.00033(34) & 0.13492 &$-$0.01956(58) &   0.00025(45) 
   \\
   0.13540 &$-$0.02474(30) &   0.00058(46) & 0.13500 &$-$0.02271(43) &$-$0.00014(34) 
\end{tabular}
\end{ruledtabular}

\label{tab:Nf3:MdM}
\vspace{-5ex}
\end{center}
\end{table}

\addtocounter{table}{-1}
\begin{table}
\begin{center}
\caption{({\it Continued})}

\begin{ruledtabular}
\begin{tabular}{ccc||ccc}
   \multicolumn{6}{c}{$\beta\!=\!6.3$}
   \\ \hline
   \multicolumn{3}{c||}{$c_{\rm SW}\!=\!1.3117$} & 
   \multicolumn{3}{c  }{$c_{\rm SW}\!=\!1.3675$} 
   \\ \hline
   $K$      & $aM$ & $a\Delta M$ &
   $K$      & $aM$ & $a\Delta M$
   \\ \hline
   0.13501 &   0.05392(55) &   0.00085(50) & 0.13501 &   0.03677(44) &   0.00097(29) 
   \\
   0.13555 &   0.03766(40) &   0.00107(32) & 0.13555 &   0.02065(30) &   0.00063(33) 
   \\
   0.13609 &   0.02152(36) &   0.00080(38) & 0.13609 &   0.00353(36) &   0.00058(32) 
   \\
   0.13664 &   0.00403(41) &   0.00175(34) & 0.13664 &$-$0.01345(53) &   0.00168(41) 
   \\ \hline
   \multicolumn{3}{c||}{$c_{\rm SW}\!=\!1.4233$} & 
   \multicolumn{3}{c  }{$c_{\rm SW}\!=\!1.4791$} 
   \\ \hline
   $K$      & $aM$ & $a\Delta M$ &
   $K$      & $aM$ & $a\Delta M$ 
   \\ \hline
   0.13446 &   0.03635(37) &   0.00070(37) & 0.13446 &   0.01834(37) &$-$0.00034(33) 
   \\
   0.13501 &   0.01915(34) &   0.00028(37) & 0.13501 &   0.00198(38) &   0.00006(30) 
   \\
   0.13555 &   0.00284(43) &   0.00092(33) & 0.13530 &$-$0.00807(57) &   0.00075(47) 
   \\
   0.13609 &$-$0.01497(43) &   0.00050(37) & 0.13555 &$-$0.01599(41) &$-$0.00039(51) 
%
   \\ \hline
   \multicolumn{6}{c}{$\beta\!=\!6.0$}
   \\ \hline
   \multicolumn{3}{c||}{$c_{\rm SW}\!=\!1.3237$} & 
   \multicolumn{3}{c  }{$c_{\rm SW}\!=\!1.3801$} 
   \\ \hline
   $K$      & $aM$ & $a\Delta M$ &
   $K$      & $aM$ & $a\Delta M$
   \\ \hline
   0.13552 &   0.07608(51) &   0.00154(39) & 0.13552 &   0.05826(50) &   0.00156(37) 
   \\
   0.13607 &   0.06056(57) &   0.00175(43) & 0.13607 &   0.04182(39) &   0.00124(44) 
   \\
   0.13661 &   0.04411(52) &   0.00120(48) & 0.13661 &   0.02400(46) &   0.00077(40) 
   \\
   0.13716 &   0.02755(53) &   0.00241(53) & 0.13716 &   0.00843(50) &   0.00259(53) 
   \\ \hline
   \multicolumn{3}{c||}{$c_{\rm SW}\!=\!1.4364$} & 
   \multicolumn{3}{c  }{$c_{\rm SW}\!=\!1.4927$} 
   \\ \hline
   $K$      & $aM$ & $a\Delta M$ &
   $K$      & $aM$ & $a\Delta M$ 
   \\ \hline
   0.13498 &   0.05492(48) &   0.00054(40) & 0.13498 &   0.03635(37) &$-$0.00060(60) 
   \\
   0.13552 &   0.03893(49) &   0.00035(54) & 0.13552 &   0.01878(41) &   0.00030(45) 
   \\
   0.13607 &   0.02221(48) &   0.00107(44) & 0.13607 &   0.00235(45) &$-$0.00005(48) 
   \\
   0.13661 &   0.00562(49) &   0.00031(50) & 0.13634 &$-$0.00692(49) &   0.00076(39) 
%
   \\ \hline
   \multicolumn{6}{c}{$\beta\!=\!5.7$}
   \\ \hline
   \multicolumn{3}{c||}{$c_{\rm SW}\!=\!1.5431$} & 
   \multicolumn{3}{c  }{$c_{\rm SW}\!=\!1.6088$} 
   \\ \hline
   $K$      & $aM$ & $a\Delta M$ &
   $K$      & $aM$ & $a\Delta M$
   \\ \hline
   0.13575 &   0.04354(59) &   0.00079(42) & 0.13541 &   0.02748(50) &$-$0.00068(47) 
   \\
   0.13644 &   0.02091(55) &$-$0.00039(51) & 0.13610 &   0.00695(130)&   0.00053(54) 
   \\
   0.13712 &$-$0.00181(57) &   0.00072(53) & 0.13644 &$-$0.00587(70) &$-$0.00087(64) 
   \\
   0.13750 &$-$0.01386(70) &   0.00007(72) & 0.13678 &$-$0.01702(60) &$-$0.00089(65) 
   \\ \hline
   \multicolumn{3}{c||}{$c_{\rm SW}\!=\!1.6745$} & 
   \multicolumn{3}{c  }{$c_{\rm SW}\!=\!1.7401$} 
   \\ \hline
   $K$      & $aM$ & $a\Delta M$ &
   $K$      & $aM$ & $a\Delta M$ 
   \\ \hline
   0.13507 &   0.01331(69) &$-$0.00081(51) & 0.13400 &   0.02229(42) &$-$0.00140(52) 
   \\
   0.13540 &   0.00278(64) &$-$0.00022(61) & 0.13450 &   0.00521(59) &$-$0.00104(39) 
   \\
   0.13575 &$-$0.00947(54) &$-$0.00078(43) & 0.13473 &$-$0.00173(44) &$-$0.00082(42) 
   \\
   0.13585 &$-$0.01124(62) &$-$0.00115(51) & 0.13500 &$-$0.01153(59) &$-$0.00172(69) 
%
   \\ \hline
   \multicolumn{6}{c}{$\beta\!=\!5.4$}
   \\ \hline
   \multicolumn{3}{c||}{$c_{\rm SW}\!=\!1.6$} & 
   \multicolumn{3}{c  }{$c_{\rm SW}\!=\!1.7$} 
   \\ \hline
   $K$      & $aM$ & $a\Delta M$ &
   $K$      & $aM$ & $a\Delta M$
   \\ \hline
   0.13750 &   0.03310(70) &   0.00015(43)   & 0.13600 &   0.03592(56) &$-$0.00031(40) 
   \\
   0.13810 &   0.01245(75) &   0.00083(52)   & 0.13630 &   0.02622(54) &   0.00032(52) 
   \\
   0.13825 &   0.00684(78) &   0.00053(62)   & 0.13660 &   0.01653(48) &   0.00030(46) 
   \\
   0.13840 &   0.00382(124) &   0.00100(49)  & 0.13720 &$-$0.00589(52) &   0.00106(57) 
   \\ \hline
   \multicolumn{3}{c||}{$c_{\rm SW}\!=\!1.8$} & 
   \multicolumn{3}{c  }{$c_{\rm SW}\!=\!1.9$} 
   \\ \hline
   $K$      & $aM$ & $a\Delta M$ &
   $K$      & $aM$ & $a\Delta M$ 
   \\ \hline
   0.13480 &   0.03073(51) &$-$0.00129(56) & 0.13330 &   0.03600(40) &$-$0.00149(40) 
   \\
   0.13520 &   0.01650(50) &$-$0.00048(49) & 0.13370 &   0.02250(34) &$-$0.00167(51) 
   \\
   0.13560 &   0.00373(52) &   0.00010(47) & 0.13410 &   0.00871(47) &$-$0.00110(40) 
   \\
   0.13580 &$-$0.00426(70) &$-$0.00062(53) & 0.13450 &$-$0.00508(51) &$-$0.00194(51) 
\end{tabular}
\end{ruledtabular}

\label{tab:Nf3:MdM2}
\vspace{-5ex}
\end{center}
\end{table}

\addtocounter{table}{-1}
\begin{table}
\begin{center}
\caption{({\it Continued})}

\begin{ruledtabular}
\begin{tabular}{ccc||ccc}
   \multicolumn{6}{c}{$\beta\!=\!5.2$}
   \\ \hline
   \multicolumn{3}{c||}{$c_{\rm SW}\!=\!1.70$} & 
   \multicolumn{3}{c  }{$c_{\rm SW}\!=\!1.85$} 
   \\ \hline
   $K$      & $aM$ & $a\Delta M$ &
   $K$      & $aM$ & $a\Delta M$
   \\ \hline
   0.1375 &   0.05191(131) &   0.00102(87)   & 0.1357 &   0.03406(87) &$-$0.00004(54) 
   \\
   0.1379 &   0.03776(89) &   0.00062(67) 	  & 0.1361 &   0.01844(73) &   0.00045(46) 
   \\
   0.1383 &   0.02311(82) &   0.00196(54) 	  & 0.1365 &   0.00287(78) &   0.00019(81) 
   \\
   0.1387 &   0.00807(123) &   0.00040(110)  & 0.1366 &   0.00037(84) &$-$0.00038(56) 
   \\ \hline
   \multicolumn{3}{c||}{$c_{\rm SW}\!=\!2.00$} & 
   \multicolumn{3}{c  }{$c_{\rm SW}\!=\!2.15$} 
   \\ \hline
   $K$      & $aM$ & $a\Delta M$ &
   $K$      & $aM$ & $a\Delta M$ 
   \\ \hline
   0.1336 &   0.02982(64) &$-$0.00161(43) & 0.1316 &   0.02568(70) &$-$0.00314(65) 
   \\
   0.1340 &   0.01467(74) &$-$0.00089(48) & 0.1320 &   0.01350(49) &$-$0.00281(46) 
   \\
   0.1344 &$-$0.00003(68) &$-$0.00219(48) & 0.1322 &   0.00364(74) &$-$0.00348(77) 
   \\
   0.1345 &$-$0.00405(69) &$-$0.00133(48) & 0.1324 &$-$0.00202(51) &$-$0.00284(60) 
\end{tabular}
\end{ruledtabular}

\label{tab:Nf3:MdM3}
\vspace{-5ex}
\end{center}
\end{table}


\begin{table}
\begin{center}
\caption{
   Fit parameters for Eq.~(\ref{eq:Nf2b52:combfit:M}) in three-flavor QCD.
}
\begin{ruledtabular}
\begin{tabular}{llllllll}
   $\beta$ & $\chi^2/{\rm dof}$ 
           & $a_M$          & $b_{M}^{(1)}$  & $b_{M}^{(2)}$
                            & $c_{M}^{(1)}$  & $c_{M}^{(2)}$ 
                            & $d_{M}$ 
\\ \hline
   12.0    & 0.96 &$-$10.9( 1.1)
                  &    2.40(28) & $-$0.125(18)
                  & $-$0.47(13) & $-$0.012(23)
                  &    0.046(16)\\
    9.6    & 1.41 &$-$13.1(1.1)
                  &    3.02(30) & $-$0.168(20)
                  & $-$0.64(14) & $-$0.047(22)
                  &    0.076(17)\\ 
    7.4    & 1.17 &$-$15.9(2.9)
                  &    3.90(81) & $-$0.233(55)
                  & $-$0.90(34) & $-$0.057(39)
                  &    0.106(46)\\
    6.8    & 1.57 &$-$24.9(4.3)
                  &    6.6(1.2)  & $-$0.430(85)
                  & $-$2.30(47)  &    0.016(53)
                  &    0.266(68)\\
    6.3    & 0.66 & $-$19.0(5.8)
                  &    5.0(1.7) & $-$0.33(12)
                  & $-$1.99(66) & $-$0.094(40)
                  &    0.262(97)\\
    6.0    & 1.19 & $-$12.2(6.6)
                  &    3.2(1.9)  & $-$0.20(13)
                  & $-$1.75(64)  & $-$0.111(40)
                  &    0.234(92)\\
    5.7    & 1.16 & $-$7.6(9.5)
                  &    1.6(3.0) & $-$0.07(23)
                  & $-$0.4(1.8) & $-$0.007(99)
                  &    0.005(280)\\
    5.4    & 0.92 & $-$27(11)
                  &    8.0(3.7) & $-$0.58(30)
                  & $-$5.0(2.7) & $-$0.19(17)
                  &    0.70(45)\\
    5.2    & 1.42 &$-$38(22)
                  &   11.9(7.4) & $-$0.91(62)
                  & $-$8.0(5.5) & $-$0.32(35)
                  &    1.17(93)
\end{tabular}
\end{ruledtabular}
\label{tab:Nf3:Mfit}
\vspace{-5ex}
\end{center}
\end{table}

\begin{table}
\begin{center}
\caption{
   Fit parameters for Eq.~(\ref{eq:Nf2b52:combfit:dM}) in three-flavor QCD.
}
\begin{ruledtabular}
\begin{tabular}{llllllll}
   $\beta$ & $\chi^2/{\rm dof}$ 
           & $a_{\Delta M}$ & $b_{\Delta M}^{(1)}$ & $b_{\Delta M}^{(2)}$
                            & $c_{\Delta M}^{(1)}$ & $c_{\Delta M}^{(2)}$ 
                            & $d_{\Delta M}$
\\ \hline
   12.0    & 0.59 &    1.6( 1.2)
                  & $-$0.41(32) &    0.025(21)
                  & $-$0.12(16) & $-$0.008(26)
                  &    0.016(19)\\
    9.6    & 0.91 &    0.6(1.3)
                  & $-$0.14(35) &    0.008(23)
                  & $-$0.14(16) & $-$0.007(25)
                  &    0.019(19)\\
    7.4    & 1.13 & $-$3.1(3.4)
                  &    0.82(93)  & $-$0.053(64)
                  &    0.09(38)  &    0.025(44)
                  & $-$0.022(51)\\
    6.8    & 0.40 &    6.2(4.4)
                  & $-$1.7(1.2)  &    0.109(84)
                  & $-$0.06(50)  & $-$0.051(49)
                  &    0.025(66)\\
    6.3    & 1.14 &    7.1(5.3)
                  & $-$2.0(1.5) &    0.14(11)
                  &    0.49(61) & $-$0.012(38)
                  & $-$0.062(91)\\
    6.0    & 1.28 &    8.7(6.3)
                  & $-$2.5(1.8)  & 0.18(13)
                  &    0.78(60)  & 0.003(37)
                  & $-$0.108(84)\\
    5.7    & 1.06 & $-$4.8(8.7)
                  &    1.5(2.7)  & $-$0.11(21)
                  & $-$0.7(1.5)  &    0.008(80)
                  &    0.09(23)\\
    5.4    & 0.53 & $-$6.9(10.8)
                  &    2.4(3.6) & $-$0.21(29)
                  & $-$2.1(2.6) & $-$0.18(16)
                  &    0.38(43)\\
    5.2    & 0.67 & $-$26(16)
                  &    8.7(5.4) & $-$0.73(45)
                  & $-$6.6(4.0) & $-$0.42(25)
                  &    1.10(67)
\end{tabular}
\end{ruledtabular}
\label{tab:Nf3:dMfit}
\vspace{-5ex}
\end{center}
\end{table}


\begin{table}
\begin{center}
\caption{
   Non-perturbative estimate of $c_{\rm SW}$ and $K_c$ 
   in three-flavor QCD
   obtained from tree-level improvement condition 
   Eq.~(\ref{eq:setup:ImpCnd}).
}
\begin{ruledtabular}
\begin{tabular}{lllllllllll}
   $\beta$ & $c_{\rm SW}$ & $K_c$ 
   \\ \hline
   12.0 & 1.1415(48)& 0.129841(21) \\
    9.6 & 1.1916(50)& 0.131321(28) \\
    7.4 & 1.316(11) & 0.133567(87) \\
    6.8 & 1.358(13) & 0.13466(12)  \\
    6.3 & 1.447(15) & 0.13539(16)  \\
    6.0 & 1.494(14) & 0.13612(16)  \\
    5.7 & 1.544(32) & 0.13706(39)  \\
    5.4 & 1.740(30) & 0.13650(41)  \\
    5.2 & 1.764(103) & 0.13789(159)
\end{tabular}
\end{ruledtabular}
\label{tab:Nf3:cSW}
\vspace{-5ex}
\end{center}
\end{table}

\clearpage


\begin{table}
\begin{center}
\caption{Quark mass $M$ and its difference $\Delta M$ in four-flavor QCD
         at $\beta\!=\!9.6$.
}

\begin{ruledtabular}
\begin{tabular}{ccc||ccc}
   \multicolumn{3}{c||}{$c_{\rm SW}\!=\!1.12990$} & 
   \multicolumn{3}{c  }{$c_{\rm SW}\!=\!1.16539$} 
   \\ \hline
   $K$      & $aM$ & $a\Delta M$ &
   $K$      & $aM$ & $a\Delta M$
   \\ \hline
   0.13043 &   0.03530(11) &   0.00111(14) & 0.13043 &   0.02955(11) &   0.00050(11) 
   \\
   0.13109 &   0.01527(12) &   0.00118(15) & 0.13109 &   0.00892(12) &   0.00068(13) 
   \\
   0.13175 &$-$0.00522(12) &   0.00110(12) & 0.13175 &$-$0.01158(11) &   0.00069(12) 
   \\
   0.13240 &$-$0.02611(13) &   0.00123(12) & 0.13240 &$-$0.03256(14) &   0.00061(13) 
   \\ \hline
   \multicolumn{3}{c||}{$c_{\rm SW}\!=\!1.20089$} & 
   \multicolumn{3}{c  }{$c_{\rm SW}\!=\!1.23638$} 
   \\ \hline
   $K$      & $aM$ & $a\Delta M$ &
   $K$      & $aM$ & $a\Delta M$ 
   \\ \hline
   0.13043 &   0.02328(12) &   0.00018(11) & 0.13043 &   0.01723(12) &$-$0.00032(14) 
   \\
   0.13109 &   0.00280(14) &   0.00028(13) & 0.13109 &$-$0.00364(12)
   &$-$0.00021(11) 
   \\
   0.13175 &$-$0.01825(10) &   0.00028(15) & 0.13175 &$-$0.02462(17) &$-$0.00035(15) 
   \\
   0.13240 &$-$0.03903(15) &   0.00004(15) & 0.13240 &$-$0.04592(14) &$-$0.00023(13) 
\end{tabular}
\end{ruledtabular}

\label{tab:Nf4:MdM}
\vspace{-5ex}
\end{center}
\end{table}

\begin{table}
\begin{center}
\caption{Quark mass $M$ and its difference $\Delta M$ in two-flavor QCD
         at $\beta\!=\!9.6$.
}

\begin{ruledtabular}
\begin{tabular}{ccc||ccc}
   \multicolumn{3}{c||}{$c_{\rm SW}\!=\!1.12990$} & 
   \multicolumn{3}{c  }{$c_{\rm SW}\!=\!1.16539$} 
   \\ \hline
   $K$      & $aM$ & $a\Delta M$ &
   $K$      & $aM$ & $a\Delta M$
   \\ \hline
   0.13043 &   0.04089(10) &   0.00127(13) & 0.13043 &   0.03508(10) &   0.00085(12) 
   \\
   0.13109 &   0.02085(10) &   0.00119(13) & 0.13109 &   0.01482(12) &   0.00053(16) 
   \\
   0.13175 &   0.00047(12) &   0.00102(16) & 0.13175 &$-$0.00585(13) &   0.00070(14) 
   \\
   0.13240 &$-$0.01998(13) &   0.00104(13) & 0.13240 &$-$0.02646(15) &   0.00068(15) 
   \\ \hline
   \multicolumn{3}{c||}{$c_{\rm SW}\!=\!1.20089$} & 
   \multicolumn{3}{c  }{$c_{\rm SW}\!=\!1.23638$} 
   \\ \hline
   $K$      & $aM$ & $a\Delta M$ &
   $K$      & $aM$ & $a\Delta M$ 
   \\ \hline
   0.13043 &   0.02884(12) &   0.00013(13) & 0.13043 &   0.02259(10) &$-$0.00021(13) 
   \\
   0.13109 &   0.00855(11) &   0.00045(14) & 0.13109 &   0.00218(10) &$-$0.00003(12) 
   \\
   0.13175 &$-$0.01225(13) &   0.00013(19) & 0.13175 &$-$0.01857(13) &$-$0.00000(13) 
   \\
   0.13240 &$-$0.03270(12) &   0.00032(22) & 0.13240 &$-$0.03927(11) &$-$0.00024(16) 
\end{tabular}
\end{ruledtabular}

\label{tab:Nf2:MdM}
\vspace{-5ex}
\end{center}
\end{table}

\begin{table}
\begin{center}
\caption{Quark mass $M$ and its difference $\Delta M$ in quenched QCD
         at $\beta\!=\!9.6$.
}

\begin{ruledtabular}
\begin{tabular}{ccc||ccc}
   \multicolumn{3}{c||}{$c_{\rm SW}\!=\!1.12990$} & 
   \multicolumn{3}{c  }{$c_{\rm SW}\!=\!1.16539$} 
   \\ \hline
   $K$      & $aM$ & $a\Delta M$ &
   $K$      & $aM$ & $a\Delta M$
   \\ \hline
   0.13043 &   0.04646(7) &   0.00141(10) & 0.13043 &   0.04059(8) &   0.00095(11) 
   \\
   0.13109 &   0.02679(8) &   0.00131(11) & 0.13109 &   0.02064(8) &   0.00080(11) 
   \\
   0.13175 &   0.00656(7) &   0.00131(14) & 0.13175 &   0.00041(7) &   0.00084(10) 
   \\
   0.13240 &$-$0.01324(8) &   0.00132(10) & 0.13240 &$-$0.01978(8) &   0.00074(11) 
   \\ \hline
   \multicolumn{3}{c||}{$c_{\rm SW}\!=\!1.20089$} & 
   \multicolumn{3}{c  }{$c_{\rm SW}\!=\!1.23638$} 
   \\ \hline
   $K$      & $aM$ & $a\Delta M$ &
   $K$      & $aM$ & $a\Delta M$ 
   \\ \hline
   0.13043 &   0.03458(9) &   0.00036(11)   & 0.13043 &   0.02840(8) &$-$0.00015(10)  
   \\
   0.13109 &   0.01451(7) &   0.00034(12) 	 & 0.13109 &   0.00818(7) &   0.00006(10)  
   \\
   0.13175 &$-$0.00580(8) &   0.00050(14) 	 & 0.13175 &$-$0.01231(8) &$-$0.00005(13)  
   \\
   0.13240 &$-$0.02607(10) &   0.00053(16)  & 0.13240 &$-$0.03247(10) &$-$0.00013(21) 
\end{tabular}
\end{ruledtabular}

\label{tab:Nf0:MdM}
\vspace{-5ex}
\end{center}
\end{table}

\begin{table}
\begin{center}
\caption{
   Fit parameters in Eq.~(\ref{eq:Nf2b52:combfit:M}) for
   $N_f\!=\!4$, $2$ and $0$.
}
\begin{ruledtabular}
\begin{tabular}{lllllllll}
   $N_f$   & $\beta$ & $\chi^2/{\rm dof}$ 
           & $a_M$          & $b_{M}^{(1)}$  & $b_{M}^{(2)}$
                            & $c_{M}^{(1)}$  & $c_{M}^{(2)}$ 
                            & $d_{M}$
\\ \hline
 4  & 9.6  & 0.96 &$-$14.7(1.3)
                  &    3.52(33) & $-$0.206(22)
                  & $-$1.14(15) & $-$0.041(25)
                  &    0.140(19)\\ 
 \hline
 2  & 9.6  & 0.48 &$-$13.5(1.2)
                  &    3.15(31) & $-$0.177(20)
                  & $-$0.64(14) & $-$0.052(23)
                  &    0.077(17)\\
 \hline
 0  & 9.6  & 1.03 &$-$11.71(80)
                  &    2.70(21) & $-$0.150(14)
                  & $-$0.80(10) & $-$0.040(16)
                  &    0.094(12)\\
\end{tabular}
\end{ruledtabular}
\label{tab:Nf024:Mfit}
\vspace{-5ex}
\end{center}
\end{table}

\begin{table}
\begin{center}
\caption{
   Fit parameters in Eq.~(\ref{eq:Nf2b52:combfit:dM}) 
   for $N_f\!=\!4$, $2$ and $0$.
}
\begin{ruledtabular}
\begin{tabular}{lllllllll}
   $N_f$   & $\beta$ & $\chi^2/{\rm dof}$ 
           & $a_{\Delta M}$ & $b_{\Delta M}^{(1)}$ & $b_{\Delta M}^{(2)}$
                            & $c_{\Delta M}^{(1)}$ & $c_{\Delta M}^{(2)}$ 
                            & $d_{\Delta M}$
\\ \hline
 4  & 9.6  & 0.42 & $-$1.2(1.3)
                  &    0.33(34) & $-$0.023(22)
                  & $-$0.14(16) &    0.015(26)
                  &    0.012(19)\\
 \hline
 2  & 9.6  & 0.80 & $-$1.0(1.4)
                  &    0.24(38) & $-$0.014(25)
                  &    0.15(17) &    0.013(28)
                  & $-$0.026(21)\\
 \hline
 0  & 9.6  & 0.53 & $-$0.3(1.2)
                  &    0.06(31) & $-$0.002(20)
                  &    0.11(14) &    0.012(23)
                  & $-$0.020(18)\\
\end{tabular}
\end{ruledtabular}
\label{tab:Nf024:dMfit}
\vspace{-5ex}
\end{center}
\end{table}

\begin{table}
\begin{center}
\caption{
   Non-perturbative estimate of $c_{\rm SW}$ and $K_c$
   for $N_f\!=\!4$, 2 and 0
   obtained from tree-level improvement condition 
   Eq.~(\ref{eq:setup:ImpCnd}).
}
\begin{ruledtabular}
\begin{tabular}{lllllllllll}
   $N_f$ & $\beta$ & $c_{\rm SW}$ & $K_c$
   \\ \hline
 4 &  9.6 & 1.1954(48)  & 0.131210(27)\\
 \hline
 2 &  9.6 & 1.2028(63)  & 0.131353(36)\\
 \hline
 0 &  9.6 & 1.2112(44)  & 0.131502(25)\\
\end{tabular}
\end{ruledtabular}
\label{tab:Nf024:cSW}
\vspace{-5ex}
\end{center}
\end{table}


\begin{table}
\begin{center}
\caption{
   Non-perturbative estimate of $c_{\rm SW}$ and $K_c$ 
   in three-flavor QCD
   obtained from one-loop level improvement condition
   Eq.~(\ref{eq:OaL:ImpCnd2}).
}
\begin{ruledtabular}
\begin{tabular}{lll}
   $\beta$ & $c_{\rm SW}$ & $K_c$ 
   \\ \hline
   12.0 & 1.1507(46) & 0.129801(20)\\
    9.6 & 1.2039(46) & 0.131252(26)\\
    7.4 & 1.336(14)  & 0.13340(12)\\
    6.8 & 1.378(12)  & 0.13447(11)\\
    6.3 & 1.470(16)  & 0.13515(17)\\
    6.0 & 1.508(17)  & 0.13596(20)\\
    5.7 & 1.569(27)  & 0.13676(32)\\
    5.4 & 1.770(26)  & 0.13609(35)\\
    5.2 & 1.811(67)  & 0.1372(10)
\end{tabular}
\end{ruledtabular}
\label{tab:FSC:cSW}
\vspace{-5ex}
\end{center}
\end{table}


\clearpage


\clearpage

\clearpage


\end{document}